# Ultrafast Electron Diffraction at Surfaces: From Nanoscale Heat Transport to Driven Phase Transitions

Michael Horn-von Hoegen[*)]

Department of Physics and Center for Nanointegration CeNIDE,
University of Duisburg-Essen, Lotharstrasse. 1, 47057 Duisburg, Germany

[*)] Corresponding author, phone +49 203 379 1438, fax +49 203 379 1555, horn-von-hoegen@uni-due.de

**Abstract**

Many fundamental processes of structural changes at surfaces occur on a pico- or femtosecond time scale. In order to study such ultra-fast processes, we have combined modern surface science techniques with fs-laser pulses in a pump-probe scheme. Reflection high energy electron diffraction (RHEED) with grazing incident electrons ensures surface sensitivity for the probing electron pulses. Utilizing the Debye-Waller effect, we studied the nanoscale heat transport from an ultrathin film through a hetero-interface or the damping of vibrational excitations in monolayer adsorbate systems on the lower ps-time scale. By means of spot profile analysis the different cooling rates of epitaxial Ge nanostructures of different size and strain state were determined. The excitation and relaxation dynamics of a driven phase transition far away from thermal equilibrium is demonstrated using the In-induced (8×2) reconstruction on Si(111). This Peierls-distorted surface charge density wave system exhibits a discontinuous phase transition at 130 K from a (8×2) insulating ground state to (4×1) metallic excited state. Upon excitation by a fs-laser pulse, this structural phase transition is non-thermally driven in only 700 fs into the excited state. A small barrier of 40 meV hinders the immediate recovery of the groundstate and the system is found in a metastable supercooled state for up to few nanoseconds.

## 1 Introduction

Many processes on surfaces occur on femto- or picosecond timescales. While changes in the electronic structure can be determined through ultra-fast photo electron emission spectroscopy, this has yet not been possible for changes in the geometric position of the atoms at the surface. Our method of choice to accomplish this task is ultra-fast electron diffraction in a pump-probe setup, as sketched in Fig. 1. The surface is excited by an ultra-short laser pulse (pump) and the transient changes in an electron diffraction pattern are recorded after a time delay $\Delta t$ with an ultra-short electron pulse (probe). For negative delays, the sample is probed prior to the excitation and the ground state is accessible. With a systematic variation of the time delay, the transient response of the surface upon excitation could be determined and a movie of diffraction patterns as a function of $\Delta t$ was recorded. This movie is shown at the bottom part of Fig. 1, with snapshots of the diffraction pattern of a 6 nm-thick epitaxial Bi film on Si prior to the excitation and after the excitation for various time delays.

Surface sensitivity is achieved using electrons with grazing incidence, i.e., in a RHEED geometry, which exhibits a scattering cross-section that is $10^4$-$10^5$ larger than x-rays. These electrons are, therefore, highly sensitive to changes of atomic positions even in a film as thin as a monolayer of adsorbate atoms [1,2]. Here, we describe the setup of such an ultra-fast electron diffraction experiment and demonstrate with a few examples the huge potential of this technique for studying transient phenomena that are also far away from equilibrium at surfaces. The practicability of such an experiment was demonstrated by the early work of Elsayed-Ali [3,4] with sub nano-second temporal resolution and also by Aeschlimann et al., who improved the temporal resolution to a few tenths of picoseconds and studied transient heating and cooling of a platinum surface [5].

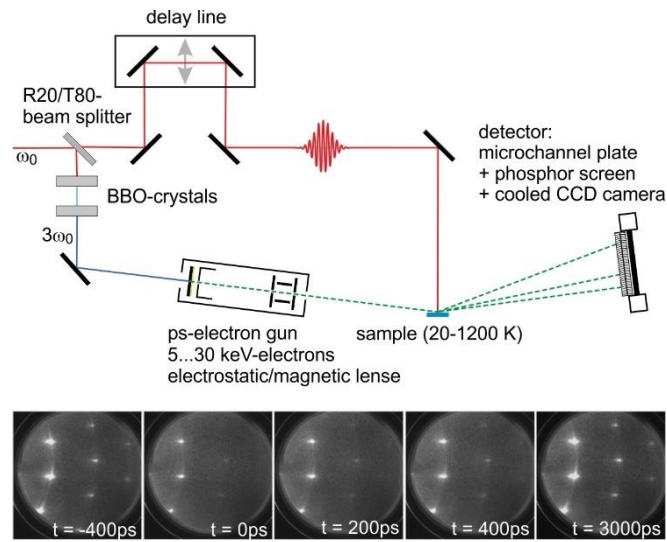

**FIG. 1.** Sketch of the pump-probe setup of the ultra-fast time resolved RHEED experiment. The sample is excited by an infrared laser pulse. Part of the initial pulse is frequency tripled and generates the ultra-short electron pulse through single electron emission in a back-illuminated transparent photocathode. The electron pulse is accelerated to an energy of 5-30 keV and subsequently diffracted at the sample surface with grazing incidence. The time delay between the optical pump and the electron probe is varied by a mechanical delay line. The series of electron diffraction patterns of a Bi(111) film on Si(111) depicts snapshots of the transient intensity drop upon excitation with the fs laser pulse and the recovery of intensity [6].

## 2 Material and methods

### 2.1 Experimental

All experiments were performed under ultra-high vacuum (UHV) conditions with a background pressure of less than $2 \times 10^{-10}$ mbar in one single UHV chamber. Beside the RHEED setup the UHV chamber was also equipped with a conventional low energy electron diffraction (LEED) instrument for sample inspection and control of sample preparation.

Silicon samples measuring $16 \times 2 \times 0.5$ mm were introduced through a load-lock system for easy sample exchange. The samples were mounted on a piezo motor driven rotatable sample stage for adjusting the azimuthal angle. The sample stage was connected to a cryostat, which allows cooling to 90 K with $LN_2$ and 20 K with He. The silicon samples could be heated up to 1400 °C by applying direct current heating. The manipulator allows motion with three degrees of freedom in translation and a second axis of rotation for the precise adjustment of the diffraction condition.

Si samples were prepared by degassing at 600 °C, followed by a short flash annealing close to the melting point, removing the native oxide. The deposition of adsorbates and the growth of epitaxial thin films was achieved *in-situ* through molecular beams from small Knudsen cell or e-beam evaporators in the same chamber [7]. Refreshing the surfaces of the samples through removal of adsorbates originating from residual gas was possible through repeated moderate re-heating with subsequent cooling.

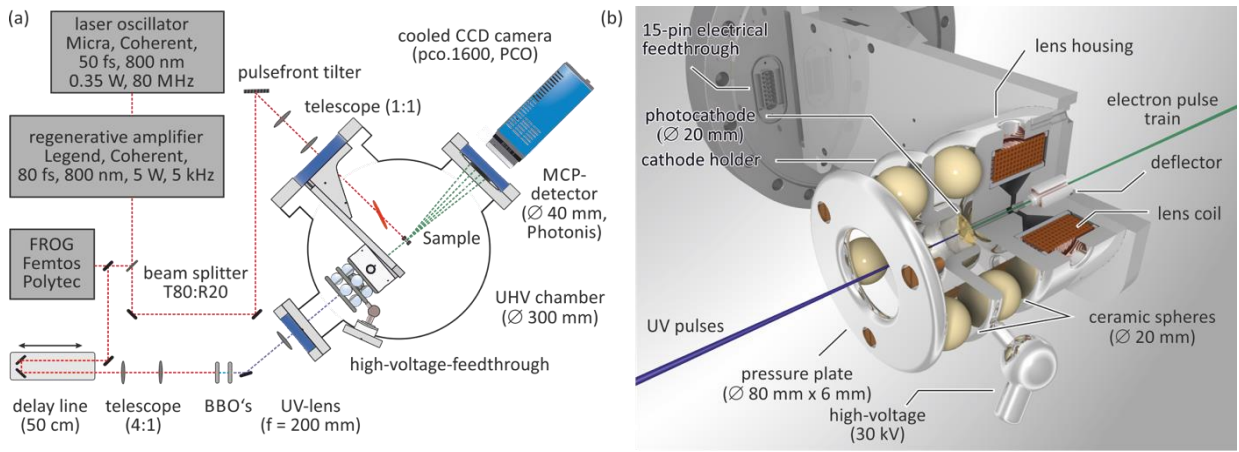

**FIG. 2.** (a) Experimental setup of the time-resolved RHEED experiment. A horizontal cut through the UHV chamber shows the fs-electron gun, the sample position, and the detector unit. The paths of the electron pulse and diffracted electrons are shown as dashed green lines. The laser system and pulse front tilting are sketched. The pump laser path is shown as dashed red line. (b) Rendered image of the third generation fs-electron gun [8]. The blue and green beam are the UV- and electron pulse, respectively. The shiny spheres are ceramic insulators to provide high voltage to the photocathode holder. Magnetic lens, magnetic deflectors, and HV-feedthrough are also shown.

The sample is excited through a regenerative titanium-sapphire laser amplifier at a repetition rate of 5 kHz, providing infrared light with a wavelength of 800 nm, i.e., a photon energy of 1.55 eV. The pulse duration is 50 fs to 80 fs with a pulse energy of 1 mJ. The pulse energy has to be high enough to provide homogenous excitation of a sample area larger than the area that is probed by the electron beam. Due to the grazing incidence of the electron beam probes together with a width of the sample of 2 mm, the pumping laser pulse should not be focused to less than 4 mm in diameter. With our setup, we achieve a laser fluence up to 14 mJ/cm$^2$.

With a beam splitter, a small part of the laser pulse is split off and frequency tripled in two barium borate crystals. The UV-pulse of 4.6 eV passes a $MgF_2$ window and generates a fs-electron pulse through single-photon photoemission from a back-illuminated photocathode [8-12]. The number of electrons in the pulse can be adjusted by the fluence of the UV-pulse through rotation of the λ/2-waveplate in front of the BBO crystals. The electrons are accelerated in a high extraction field of 7.5 kV/mm between the cathode and the anode to a kinetic energy of 30 keV. Space charge broadening and vacuum dispersion of the electron pulse requires the use of fast electrons, here travelling at 1/3 of the speed of light. The larger the extraction field and the higher the final electron energy, the smaller the temporal broadening due to the initial energy spread $\Delta E$ due to photoemission and space charge broadening effects [13]. The initial energy spread is only $\Delta E = 0.1$ eV due to the use of a 10 nm-thick gold-photocathode, where the workfunction of the Au film nicely matches the energy of the UV photons (3hν = 4.65 eV) [11].

Subsequently the electron pulse is focused by a magnetic lens located between anode and sample. At the exit of the magnetic lens, a x-y-deflector has been integrated which is composed of two split pair coils allowing us to deflect the electron beam can by ± 4° in order to change the electrons incident angle on the sample. The electron pulses are scattered under a grazing angle of 1° to 6° at the sample which is placed 50 mm beyond the guns exit aperture. The diffracted electrons are amplified through a micro-channel-plate and recorded by a cooled CCD camera.

Along the Laue-circles the width of the diffraction spots amounts to 9 % of the Brillouin-zone which is equal to 0.17 Å$^{-1}$. Perpendicular to the Laue-circles, here the 3/7 circle, the FWHM is only 0.65 % of the Brillouin-zone which is equal to 0.012 Å$^{-1}$. The transfer width is obtained by the reciprocal of the k-space resolution and amounts to $w_{trans,\parallel} = 50$ Å and $w_{trans,\perp} = 550$ Å along and perpendicular to the Laue-circle, respectively. This large difference in the resolving power is characteristic for RHEED: the direction along the electron beam path is the high-resolution direction. This high transferwidth enables ultrafast spot profile analysis as will be demonstrated in chapter Ge-hutclusters.

The grazing incidence of the electrons causes severe velocity mismatch with the laser pump pulse under normal incidence [velocity mismatch]. Though electrons of 30 keV travel already at a speed of 1/3 of light they still need 20 ps to traverse a sample of a typical width of 2 mm. Over this long time the transient intensity changes in the RHEED pattern are averaged which is disastrous for the temporal resolution! Tilting the pump pulse intensity fronts by 71° with respect to their propagation direction a constant time delay between pump and probe pulse can be achieved [14,15]. The pump pulse front tilting is achieved by first order back diffraction through a sinusoidal grating in (almost) Littrow geometry. Using a 1:1 telescope the grating is imaged onto the sample. As a result, we obtain a tilted pump-pulse front at the sample with the desired tilting angle of 71° [16]. Now the resulting width of the overall temporal response function is ultimately given by electron and laser pulse widths.

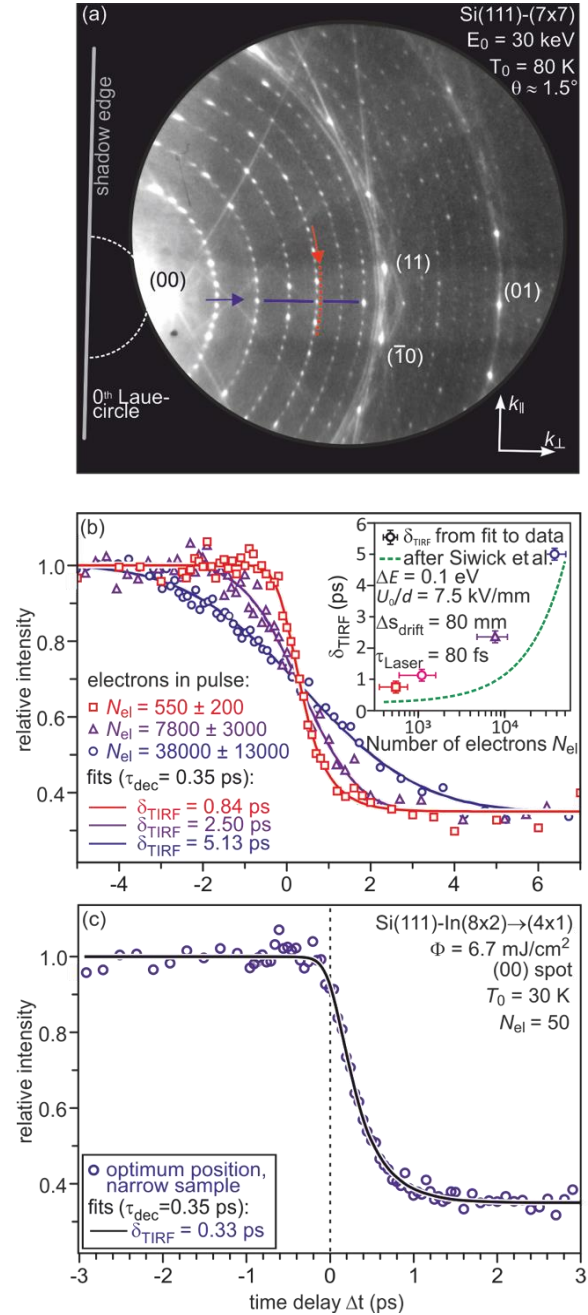

**FIG. 3.** (a) shows a diffraction pattern obtained a 80 K from the bare Si(111) surface with its inherent (7×7) reconstruction. It is characterized for a large number of sharp diffraction spots and a low back-ground intensity. In RHEED the diffraction spots are organized on so-called Laue-circles which are centered at the shadow edge beneath the (00)-spot, i.e., the specular reflected spot [17]. (b) Temporal instrumental response function for various numbers of electrons per pulse. The inset shows a comparison of the FWHM with a simulation by Siwick et al. [13]. (c) Highest possible temporal resolution of FWHM = 330 fs for a narrow sample and low number of electrons per pulse [8].

An upper limit for the temporal resolution of our RHEED experiment has been determined from the transient changes of spot intensity during the structural response of an optically driven phase transition in the Si(111)-(8×2)↔(4×1) surface CDW system (we refer to chapter Indiumwires). In Fig. 3(b) the number of electrons has been varied from 50 to 38,000 electrons per pulse. For pulses with a high electron number $N_{el}$ the temporal response becomes sigmoidal. Reducing $N_{el}$ leads to an asymmetric temporal behavior around delay zero. By fitting the above function to

- 5 -

the data, the FWHM of the temporal instrumental response function is extracted and displayed in Fig. 3(b) with a logarithmic scale of $N_{el}$. The dashed green line in the inset of Fig. 3(b) depicts the result obtained by an analytic model for electron packet propagation developed by Siwick *et.al.* [13]. The ultimate temporal resolution of 330 fs has been achieved for 50 electrons per pulse which is only slightly larger than the theoretical achievable temporal resolution of 275 fs [8].

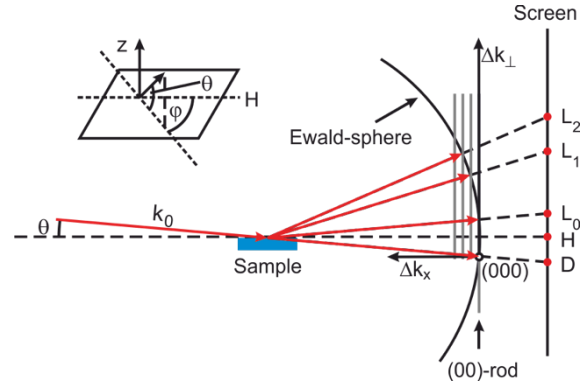

**FIG. 4.** RHEED geometry. The electron beam is incident with a grazing angle θ. The position of the diffraction spots on the screen is determined by the intersection of the lattice rods with the Ewald sphere.

The necessity of using electrons with high energy prohibits a normal incidence of the electrons for the diffraction experiment. With a dominant forward scattering and a mean free path of the order of ~5 nm, the electrons are no longer surface sensitive. The large vertical momentum transfer for backscattering under normal incidence at such high electron energies $E = 30$ keV would give rise to a huge Debye-Waller effect, i.e., would render a diffraction pattern impossible.

Surface sensitivity is achieved using a grazing incidence angle between 1° – 6° in a RHEED geometry. Then the vertical momentum transfer:

$$|\mathbf{k}_\perp| = 4\pi \sin\theta / \lambda_{el} \qquad (1)$$

is of the same order as in a typical LEED setup, i.e., for $E = 30$ keV energy at a typical grazing angle $\theta = 2.5°$ we obtain $|\mathbf{k}_\perp| \cong 7.8$ Å$^{-1}$, which corresponds to $E \cong 57$ eV at normal incidence in a LEED experiment.

The RHEED pattern is amplified with a multichannel plate and recorded from a phosphorus screen with a cooled CCD camera. To obtain a movie of RHEED patterns with varying time delays Δt, the arrival time of the pump laser pulse is changed through an mechanical delay line with a length of $2 \times 50$ cm. This accounts for a possible range for the time delay of $\Delta t = -300$ ps … 3000 ps.

Averaging and recording a typical RHEED pattern takes 10 seconds. In order to account for drift effects and slowly varying intensity fluctuations of the laser, each diffraction pattern is normalized by a pattern recorded without laser excitation of the sample. Obtaining an entire movie with 400 frames takes about two hours.

## 2.2 Spot Position in Diffraction

In general, the position of the diffraction spots is determined by the intersection of the Ewald sphere with the vertical lattice rods [17,18] originating from the 2-dimensional surface lattice as sketched in Fig. 4. Because the Ewald sphere and the reciprocal lattice rods intersect under grazing angle, the diffraction pattern is strongly distorted as compared to LEED. The RHEED spots are located on the so-called Laue circles.

In a stationary experiment, the position of all spots - except for the (00) spot - varies as a function of temperature, thus reflecting the thermal expansion of the lattice. Such a shift of spot positions, however, could not been observed in an ultra-fast RHEED experiment on a flat surface. As already stated, the spot position is determined by the intersection of reciprocal lattice rods with the Ewald sphere. The location of the rods in reciprocal space is solely determined by the lateral lattice parameter of the surface. A sample with a macroscopic width $b$ can expand laterally not faster than the speed of sound $c_s$. A change of the lateral lattice parameter $a_\parallel$ may then be expected for times $t \geq b/c_s$. For typical values of $b = 1$ mm and $c_s = 10^4$ m/s, the lateral expansion of the lattice will take at least 100 ns, i.e., much longer than any delay time that can be accomplished by an optical delay stage of a reasonable length! The sample needs an extremely long time to react on the excitation by macroscopic lateral thermal expansion.

It is important to note that an expansion of the vertical lattice parameter, i.e., the distance between an adsorbate layer and the substrate, may occur much faster than picoseconds. A variation of this parameter, however, has no influence on the spot position in RHEED (or LEED). Of course, the intensity of the spots will be affected due to a change of the dynamic form factor, which is caused by the change of the unit cell geometry [17]. A direct measurement of the layer separation from transient shifts of spot position is not possible with RHEED!

Any shifts of spot positions arise either from the lateral expansion of small islands of a very rough surface or by artifacts arising from space charge deflection in front of the sample [19-22]. This space charge is generated by nonlinear photoemission induced by the intensely pumping laser pulse. Such nonlinear effects are greatly enhanced for rough surfaces [23] and higher laser fluences. In our experiments, however, we never observed transient spot shifts up to the highest incident laser fluences of~ 14 mJ/cm$^2$.

## 3 Results

In the following sections, we will focus on a few examples to demonstrate how versatile this technique is for studies of dynamic processes on the pico- and femtosecond timescale in surface science and nanoscale physics.

### 3.1 Debye-Waller Effect of/in/during/for Thin Film Cooling: Bi films on Si substrates

One of the simplest experiments to demonstrate the proof of principle of time-resolved RHEED is the observation of transient heating and subsequent cooling of a single crystalline film on a substrate upon fs-laser excitation. The decrease of spot intensity is interpreted by means of the Debye-Waller effect as a transient rise of temperature [24,25]. For an isotropic thermal motion of the atoms, the decrease of intensity is determined by the averaged square of the vibrational amplitude $\langle \mathbf{u}^2 \rangle$ multiplied with the square of the momentum transfer $\mathbf{k}^2$:

$$I = e^{-2M}, \quad 2M = \frac{1}{3}|\mathbf{k}|^2 \langle \mathbf{u}^2 \rangle. \tag{2}$$

The square of the vibrational amplitude $\langle \mathbf{u}^2 \rangle$ can be expressed by the temperature $T$, the mass of the atoms $m$ and the surface Debye temperature $\Theta_D$:

$$\langle u^2 \rangle = \frac{3\hbar^2 T}{m k_B \Theta_D^2} \tag{3}$$

applying the Einstein model with independent harmonic oscilaltors. As the vibrational amplitude $\mathbf{u}$ is inversely proportional to $\Theta_D$, we expect higher sensitivity for surfaces with low $\Theta_D$. Therefore, we used a thin epitaxial Bi films that covered both the surfaces from Si(111)-(7×7) and Si(001)-(2×1) substrates as sketched in Fig. 5(a). The LEED patterns prior to and after growth of thin Bi(111) films are shown in the panels from Fig. 5(b). The films have been grown in-situ.

The surface Debye temperature $\Theta_{D,Bi}$ of the Bi(111) film has been determined from a stationary experiment where the sample was slowly heated from 90 K to 320 K. The decrease of intensity of the (00) spot is plotted in Fig. 5(c). From the fit with an exponential function, we obtain $\Theta_D = 47$ K for the surface Debye temperature using the Einstein model.

The fs-laser excitation with a fluence of 1.3 mJ/cm$^2$ at a sample temperature of $T_0 = 80$ K results in a pronounced intensity drop of more than 60 % as shown in Fig. 5(d). Applying the surface Debye temperature of $\Theta_{D,surf} = 47$ K and a vertical momentum transfer of $\Delta k_\perp = 6.7$ Å$^{-1}$ to Eqs. (2) and (3), this converts into a sudden jump in the surface lattice temperature of $\Delta T = 120$ K up to $T_0 = 200$ K. The recovery of the ground state occurs by an exponential cooling with a time constant of $\tau_{cool} = 965$ ps which is determined through the thermal boundary conductance $G = 13.4$ MW/m$^2$K of the Bi-Si interface.

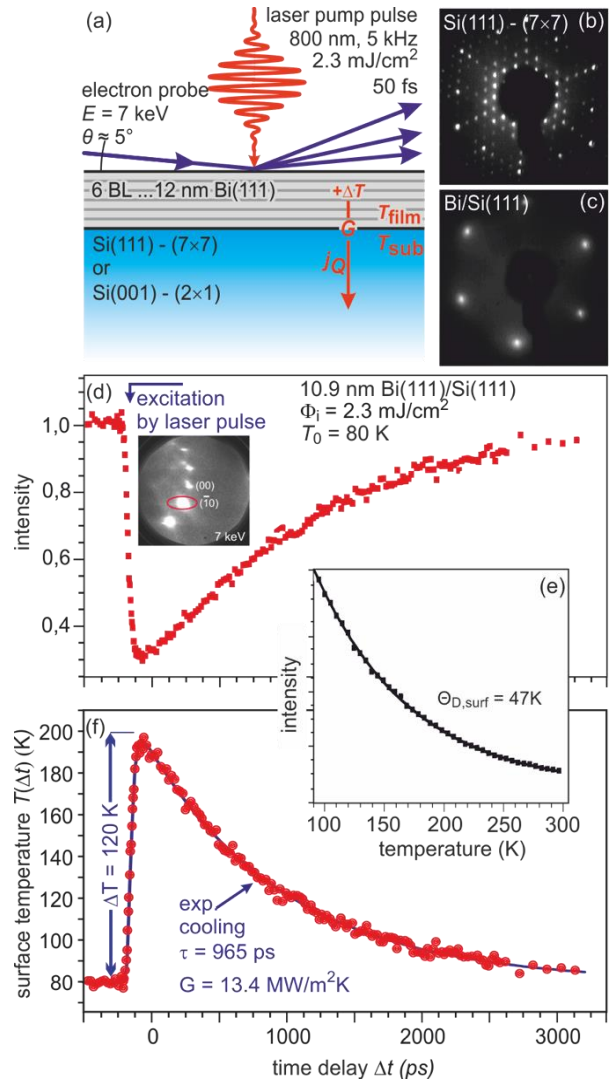

**FIG. 5.** (a) Sketch of Bi-Si heterosystem. (b) LEED patterns of Si (111) and (001) substrates prior to and after growth of ultrathin Bi(111) films. (c) The intensity of the (00) spot is plotted as function of sample temperature during quasi-stationary heating. The intensity decrease is caused by the Debye-Waller effect. From the exponential slope a surface Debye temperature $\Theta_D = 47$ K is derived. (d) The transient intensity drop of the (00) spot upon excitation with a fs-laser pulse at a fluence 2.3 mJ/cm$^2$ at a sample temperature of $T_0 = 80$ K exhibits an exponential recovery with a time constant of $\tau \cong 1000$ ps. (e) Using the stationary Debye-Waller curve taken under the very same diffraction conditions allows the direct conversion of the intensity drop to a transient temperature rise of $\Delta T = 120$ K. The exponential cooling of the Bi film with $\tau_{cool} = 965$ ps is determined through the thermal boundary conductance $G = 13.4$ MW/m$^2$K of the Bi-Si interface.

Such a slow cooling rate is described in the framework of well-established models for heat transport: the acoustic mismatch model (AMM) and the diffuse mismatch model (DMM). The discontinuity of the elastic properties at the interface between the Bi(111) film and the Si(111) substrate gives rise to an additional thermal resistance first described by Kapitza [13]. Under these conditions the slow cooling of the Bi film can be described within the DMM [26-28]. The energy is carried by phonons that were diffusively scattered at the Bi-Si interface. Applying Fermis Golden Rule the density of phonon states determines the final state, i.e., the transmission probability from Bi film into Si substrate. Due to the much lower Debye temperature of Bi in comparison with Si most of the phonons remains in the Bi film and only 1 of 100 is transmitted to the Si substrate. This results in a drastic slowdown in the cooling of the Bi film [29,30].

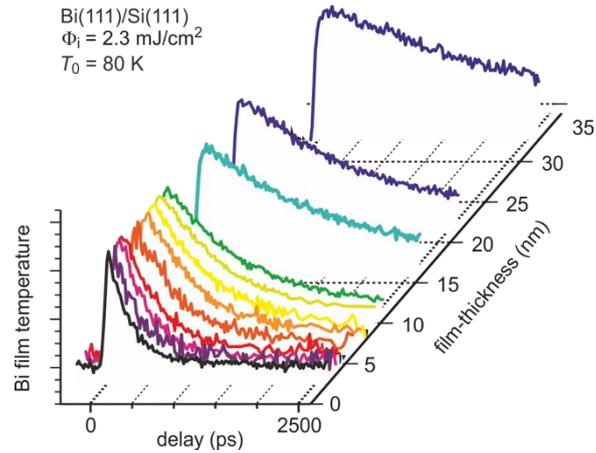

**FIG. 6.** Evolution of temperature of Bi films with various thicknesses on Si(111) upon impulsive excitation at t = 0. The cooling time constant increases with film thickness. A mono-exponential recovery to the groundstate is observed.

The Bi film thickness dependence of the cooling time constant $\tau_{cool}$ gives insight into finite size effects in nanoscale heat transfer from ultrathin films across interfaces towards substrates. The evolution of transient temperature for such a thickness series is shown in Fig. 6 for various Bi(111) films epitaxially grown on Si(111) for 2.6 nm ≤ $d_{Bi}$ ≤ 32 nm. The large variation of cooling time $\tau_{cool}$ for different film thicknesses is obvious and plotted in Fig. 7 as function of film thickness $d_{Bi}$ for Bi(111) films grown both on Si(111) and Si(001) substrates.

Figures 7(a) and 7(b) shows the behavior the Si(111) substrate in comparison with predictions from DMM. We observe the expected linear dependence of $\tau_{cool} \sim d_{Bi}$. In contrast, Figs. 7(c) and 7(d) shows a saturation of the cooling rate $\tau_{cool}$ for Bi films with $d_{Bi}$ < 6 nm. We attribute this astonishing deviation from the linear slope to a finite size effect which occurs for the AMM for films thinner than half of the mean free path of the phonons which for Bi is 12 nm at 80 K. In the framework of AMM the thermal energy is carried by phonons that were treated as elastic waves, which undergo reflection, transmission, and refraction at the interface. Because of the large difference in the velocity of sound between the film and substrate, e.g., Pb and Si, most of the phonons are trapped in the film due to total internal reflection, following Snell's law. These phonons can never escape the Bi film but may only be scattered through Normal and Umklapp-processes into the critical cone of angles α which are refracted at the interface and transmitted to the Si substrate as is sketched in Fig. 7(d), i.e., α < $\alpha_{crit}$.

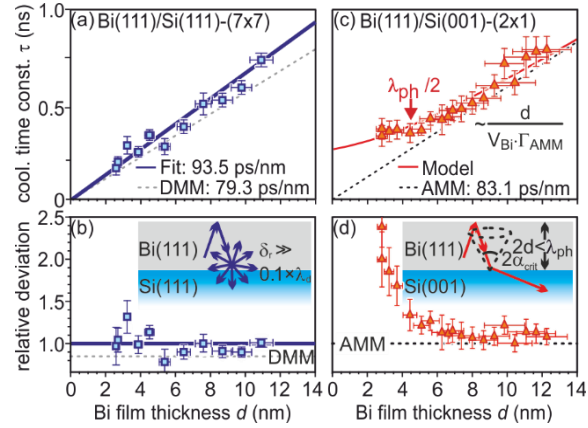

**FIG. 7.** Thickness and substrate dependence of the cooling process. (a) Cooling time constant $\tau_{cool}$ as function of film thickness d of epitaxial Bi(111) films grown on Si(111). A linear dependence of $\tau_{cool}$ with $d$ is observed. (b) Relative deviation of $\tau_{cool}$ from the linear fit shown in (a). The inset depicts the diffuse phonon scattering at the interface as expected from DMM. (c) Cooling time constant $\tau_{cool}$ of epitaxial Bi(111) films grown on Si(001). A deviation from the linear dependence of $\tau_{cool}$ with $d$ is obvious for $d < 6$ nm. (d) Relative deviation of $\tau_{cool}$ from the linear behavior which is shown as dashed line in (c). The significant deviation below 6 nm is well described by our non-equipartition model (solid red line). The inset depicts the ballistic phonon transmission with refraction and reflection at the interface as expected from AMM. The critical cone of total internal reflection is sketched by dashed lines. In all experiments the incident fluence of the 800 nm, 50 fs laser pulses was set to 2.3 mJ/cm$^2$ and the sample base temperature $T_0 = 80$ K.

### 3.2 Spot Profile Analysis of Cooling of Nanostructures: Ge clusters on Si(001)

Here, we employ nanoscale heat transport from self-organized Germanium (Ge) nanostructure into a Silicon (Si) substrate [31] to demonstrate the capabilities of tr-RHEED to observe various transient processes in parallel from the same diffraction experiment. The standard techniques for the determination of the transient temperature evolution of thin films are ultrafast optical methods like time-domain thermoreflectance (TDTR).[32,33] Due to the low scattering cross section of light with matter, they are, however, usually restricted to thicker films. Furthermore, the measured transient change of reflectivity is an integral response of the entire probed sample surface. Thus, it is not possible to distinguish between the transient response of different nanoscale structures, e.g., self-organized clusters of different dimensions which may simultaneously be present at the sample. Here, we demonstrate in detail how time-resolved spot profile analysis in electron diffraction can be used to distinguish between the transient contributions from three different cluster types upon impulsive excitation by an intense fs laser pulse.

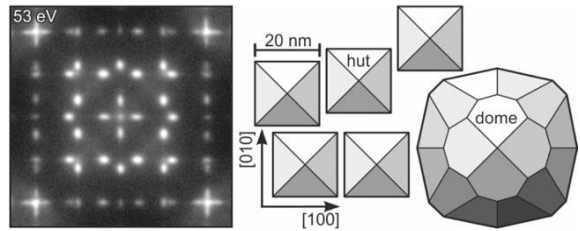

**FIG. 8**. Illustration of the two different cluster types: hut clusters, composed of four {105}-type facets and larger dome clusters with a complex faceting. The LEED pattern reflects the diffraction from the four {105}-type facets of the hut clusters.

The self-organized growth of Ge on Si(001) was utilized to prepare a sample with well defined epitaxial hut-, dome- and relaxed Ge clusters which were used as a model system. At typical growth temperatures between 400 °C and 700 °C, islanding of Ge is observed after the formation of a thin wetting layer, i.e., the so-called Stranski-Krastanov growth mode. [34] Initially, a pseudomorphous Ge film grows in a layer-by-layer fashion with the lattice constant of the underlying Si substrate. The strain energy increases with film thickness. This wetting layer becomes unstable for thicknesses of more than three monolayers of Ge (1 ML = $6.24 \times 10^{14}$ atoms/cm$^2$). With increasing coverage, the formation of a metastable phase, so-called hut clusters, is observed, which are explained as a first step on the kinetic pathway from layer-by-layer growth to islanding. [35-37] Hut clusters are composed of four {105}-type facets, which form an angle of 11.3° with the {001} Si substrate surface plane. This leads to a rectangular or square shape, as sketched in Fig. 8. Typical dimensions are 23 nm width and a height of only 2.3 nm. [38,39] As soon as the whole surface is covered with hut clusters, which should typically be the case after about 6 ML—a transformation to the next step larger islands is observed. [40-42] The formation of steeper facets allows for a more efficient reduction of lattice mismatch induced strain than in the case of huts. [42-44] With respect to their special shape, these larger clusters are usually denoted as dome clusters. Due to their complex faceting structure, their base area exhibits an approximately round shape with a diameter of 50–60 nm at a height of 5–6 nm, as sketched in Fig. 8. Both cluster types are free of lattice mismatch relieving defects and dislocations. The build-up of strain and the subsequent strain relief by islanding is the driving force for this self-organized and kinetically self-limited formation of clusters. For this reason their size distribution is very narrow. [45-48] The strain is reduced during the different transition states from layers to huts and then to domes. Hut clusters are still lateral compressed and the vertical layer distance is significantly expanded by tetragonal distortion. [42,49] In the case of dome clusters, relaxation towards the Ge bulk lattice constant is more efficient for the dome clusters. [50] Finally, the generation of defects and dislocations accommodates the lattice mismatch and causes the formation of large and fully relaxed 3D islands. Such islands do not grow any longer in a self-organized and kinetically self-limited way. Consequently, those relaxed clusters exhibit a very broad size distribution.

Figure 9(a) shows the RHEED diffraction pattern of the initial bare Si(001) surface prior to deposition. Electrons were incident along the [110] direction. The grazing incidence of the electrons leads to a vertical penetration depth of a few Angstroms only. A series of intense spots show up located on the zero order Laue circle, i.e., clear indication for diffraction from an atomically flat, single crystalline surface.

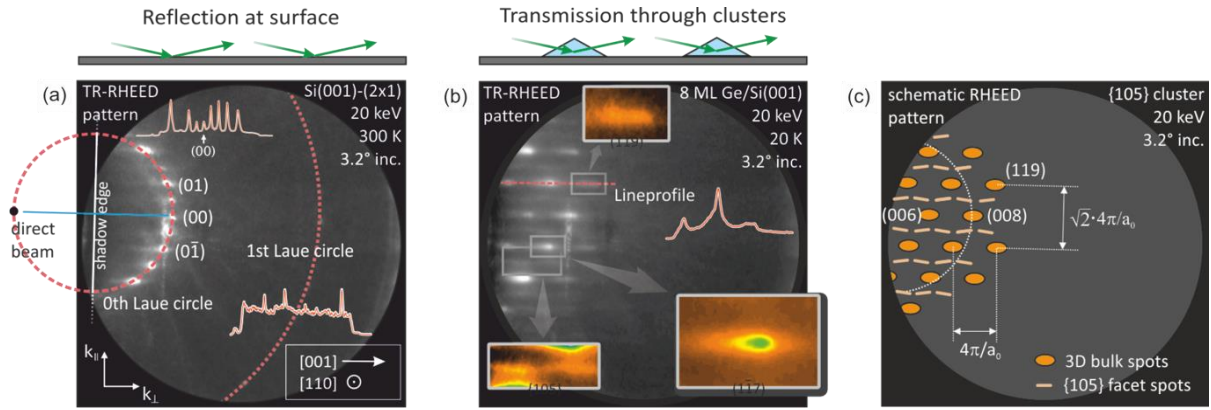

**FIG. 9.** Electron diffraction patterns of (a) the bare Si(001) surface prior to deposition and (b) after deposition of 8 ML Ge, grown at 550 °C. The diffraction geometry was in both cases the same, i.e., the incidence angle was 3.2° at an electron energy of 20 keV. Fig. (c) shows a schematic diffraction pattern of a surface covered with {105} faceted hut clusters under the same diffraction conditions. The presence of regularly ordered spots instead of a circular arrangement indicates that the diffraction happens in transmission.

After deposition of 8 ML of Ge, a sample with a high coverage of hut clusters ($8\times10^{10}$ cm$^{-2}$) and dome clusters ($2\times10^9$ cm$^{-2}$) was prepared. The change of morphology directly affects the electron diffraction pattern, as can be seen in Fig. 9(b). The typical circular arrangement of the spots for reflective diffraction from a flat surface has changed into a periodic ordered arrangement of equidistant spots which is indicative for diffraction in transmission geometry from single crystalline or epitaxial structures. This is in contrast to observations of Debye-Scherrer rings arising from diffraction from disordered, polycrystalline nanoparticles. [51] Because the grazing angle of incidence of the electrons of 3.2° is much lower than the hut cluster facet angle of 11.3°, electrons cannot be diffracted in reflection geometry from the {105} facets with orientations along the incident electron beam. Instead, they have undergone diffraction in transmission through the clusters. Thus, the diffraction pattern is described by a cut through a reciprocal diamond lattice in [001] direction. As depicted in the schematic diffraction pattern in Fig. 9(c), the distance of the spots is then given by $4\pi/a_0$ in [001] direction and in [1-10] direction by $\sqrt{2}\cdot 4\pi/a_0$, respectively, and $a_0$ the size of the cubic unit cell.

With the knowledge of the position of the (00) spot of the Si(001) substrate (see Fig. 9(a)) and the grazing angle of incidence of 3.2°, we can assign each spot to a Bragg reflection for a diamond lattice. Kinematically forbidden spots are described in terms of double diffraction, which is a prevalent effect in RHEED. [17,52] Figure 9(c) also shows the original position of the zero order Laue circle. Thus all Bragg reflections in the vicinity of this circle have a very low distance to the Ewald sphere and hence the corresponding spots exhibit a high diffraction intensity which is clearly visible in Fig. 9(b).

In addition to the transmission spots, we also expect reflection spots from those {105} facets oriented perpendicular to the incident electron beam as reported by Aumann et al. [53] Each of the transmission spots is then accompanied by four {105} facet surface diffraction spots facing towards the transmission spot. These facet spots are located between the transmission spots and can clearly be identified in the contrast enhanced inset of Fig. 9.

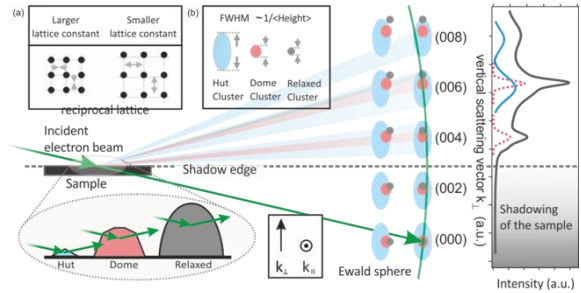

**FIG. 10.** Sketch of transmission diffraction geometry and Bragg conditions in reciprocal space for hut- (blue), dome- (red), and relaxed clusters (grey). The Ewald sphere is plotted in green. The Bragg conditions are broadened due to finite size effects and are shifted vertically and laterally due to variations of strain state. For simplicity a simple cubic lattice is used. (b) The spot profile depicts the variation of diffraction intensity of the (117) and (119) spots along $k_\perp$ direction (see also red dashed line in Fig. 9(b)). The profile can be fitted by two broad and narrow Gaussians assigned to diffraction from huts and domes, respectively.

In order to distinguish the contributions from the different cluster types to the diffraction pattern, we applied spot profile analysis. The right panel of Fig. 10 shows a line-profile along the marked red line in Fig. 9(b). The intensity is plotted as function of $k_\perp$ along the direction through the (117) and the (119) spot. The experimental data can be described by a pair of two Gaussians of different widths, positions, and intensities. The broadening and shift of spots is explained in terms of different size and strain state of the huts, domes, and relaxed clusters as is schematically depicted in the left panel of Fig. 10.

During diffraction, the finite size of the clusters effectively acts as a slit with finite width, i.e., the height, width, and length of the clusters. Thus, the diffraction spots of the smallest structures, i.e., the hut clusters, should exhibit a significant broadening in reciprocal space due to finite size effects. This broadening is most effective along the vertical momentum transfer $k_\perp$ as the smallest dimension of the huts is their height, as sketched in the inset of Fig. 10(b). For the dome clusters, we expect narrow spots because they exhibit a height which is 3-4 times larger than that of the huts. We therefore assign the broadened Gaussians (blue dotted line in Fig. 10(b)) to the 2 nm high hut clusters. The FWHM (full width at half maximum) of vertical momentum of 1.0 Å$^{-1}$ is slightly larger than the value of 0.62 Å$^{-1}$, which is expected for diffraction from the entire hut cluster, i.e., assuming infinite mean free path of the electrons, kinematic approximation, and volume weighted hut cluster profile in the direction parallel to the diffraction vector. The narrow Gaussian (red dashed line) originate from the higher and larger dome clusters.

The different strain states of the three cluster types result in different positions of the corresponding diffraction spots. Hut clusters are coherent to the Si substrate and exhibit a 4 % increased layer separation by tetragonal distortion.[41] This vertical strain causes a significant shift of the corresponding spots to lower values of $\Delta k_\perp = 62$ Å$^{-1}$, i.e., towards the (000) Bragg condition or the shadow edge, respectively. From the observed shift $\Delta k_\perp$ of the position of the broad Gaussian to lower values of $\Delta k_\perp$ we conclude an increased layer separation of $\Delta d/d = 2.6\%$ of the hut clusters with respect to the dome clusters since the apex of the dome clusters is more relaxed than hut clusters. [41,43] The lateral compression of the hut clusters towards the Si lattice parameter can also be observed through a shift of the spot positions of the broad Gaussian away from the (00) rod to larger values of $k_\|$, as sketched in Fig. 10.

Up to now, we have used the width and the position of the diffraction spots to distinguish between contributions from huts and domes. The intensities of these peaks are determined by the dynamical structure factor $F_{hkl}$ and the density of the clusters on the sample, i.e., the diffraction volume, which both are not well-defined quantities. Last but not least, also the Laue condition $\Delta \mathbf{k} = \mathbf{G}$ affects the relative intensities and may be used to distinguish between huts and domes. The Laue condition is fulfilled when reciprocal lattice points or lattice rods are intersected by the Ewald sphere, i.e., for spots on the zero order Laue circle or in its direct vicinity as sketched in Fig. 10. In case that the reciprocal lattice points are broadened (due to finite size effects), this condition becomes relaxed. Therefore, diffraction from hut clusters (broad light blue spots in Fig. 10) is still observed for larger vertical momentum transfer, where the dome clusters (light pink spots) no longer contribute to the diffraction pattern because their reciprocal lattice spot is no longer intersected by the Ewald sphere (green line). Such a case is observed for the (119) spot, where the relative intensity of the narrow Gaussian is much smaller than for the (117) spot which is located close to the zero order Laue circle. Thus, all spots on the right-hand side of the diffraction pattern in Fig. 9(b) arise almost solely from hut clusters.

The heating of the Ge clusters upon impulsive fs laser excitation and subsequent cooling to the substrate are determined from the transient intensity drop which is shown in Fig. 11 for different diffraction spots. Figure 11(a) shows the transient intensity of the (117) spot as function of time delay $\Delta t$ between the pump laser pulse and the probe electron pulse. The initial drop at the temporal overlap $\Delta t = 0$ reflects the heating from 25 K to 150 K. Applying a stationary intensity vs. sample temperature curve for calibration of the Debye-Waller effect we were able to determine the transient temperature evolution and the maximum temperature rise. upon fs laser excitation which is known to happen on a picosecond timescale.[54] Here, the observed timescale of 25 ps is determined by the temporal response function of the RHEED setup, which was at that time still dominated by the velocity mismatch between pumping laser pulse and probing electron pulse.[55] The recovery of the spot intensity occurs on a slower timescale and reflects the cooling of the Ge hut clusters through heat transfer to the Si substrate. Applying the above described spot profile analysis, the contributions from hut-, dome-, and relaxed clusters could be discriminated from the profile of the (117) spot. The solid lines are fits to the data assuming an exponential recovery of intensity.

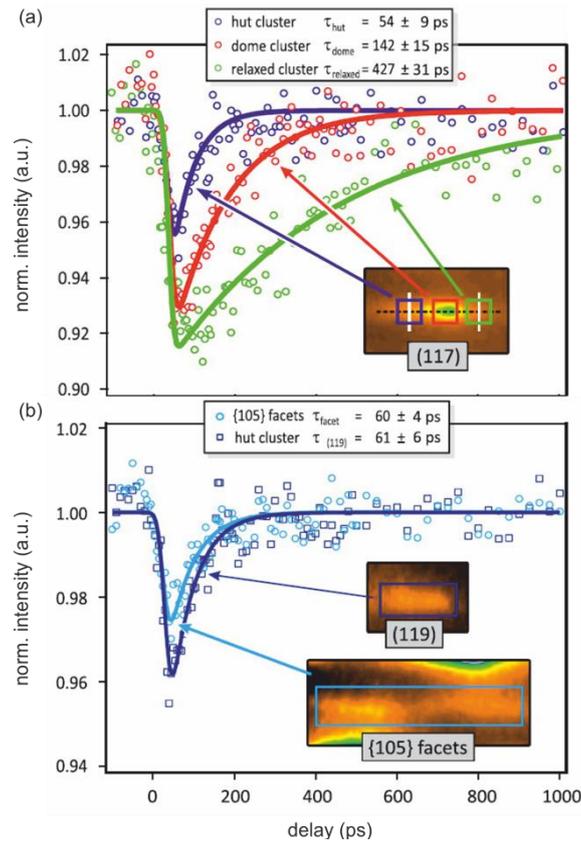

**FIG. 11.** Transient diffraction intensities as function of time delay $\Delta t$ (a) different positions of the (117) spot exhibit three different recovery time constants which are assigned to diffraction from hut-, dome-, and relaxed clusters. (b) The transient intensity of the (119) spot and the {105} facet spots originate from hut clusters. Experimental data were fitted by a convolution of monoexponential decay with a temporal response function of 25 ps width.

We observe three distinct different recovery time constants of $\tau_{hut}$ = 54 ps, $\tau_{dome}$ = 140 ps, and $\tau_{relax}$ = 430 ps, which were assigned to hut-, dome-, and relaxed clusters, respectively. The results from the transient spot profile analysis are supported by an independent analysis of the (119) spot and the {105} facet spots, which are shown in Fig. 11(b). The intensity of the (119) spot (open squares) is dominated by diffraction from hut clusters and reveals a recovery with $\tau_{(119)}$ = 61 ps. The {105} facet spots arise solely from the hut clusters and recover on a timescale of $\tau_{facet}$ = 60 ps. All three recovery time constants for the hut clusters are in good agreement, and we obtain an average cooling time constant of $\tau_{hut}$ = 58 ps.

The slower recovery time constants for the dome- and relaxed clusters were additionally confirmed by an analysis of the temporal evolution of the entire diffraction pattern, i.e., a so called lifetime map. Each pixel of the temporal series of diffraction patterns is fitted by an exponential recovery function like that used in Fig. 11. Figure 12(a) displays the recovery time constant using color coding. Blue indicates diffracted intensity with fast recovery time constant up to 100 ps, red—recovery times about 250 ps, while green indicates slow time constant of 400 to 500 ps. Black areas indicate background intensity where the signal to noise ratio was insufficient to provide a proper fit. The blue to red asymmetry of all spots—i.e., blue is on the left, red in the middle, and green is on the right wing of all spots—support the spot profile analysis we have performed before. This is shown more clearly in the inset of Fig. 12(a) showing the (117) spot.

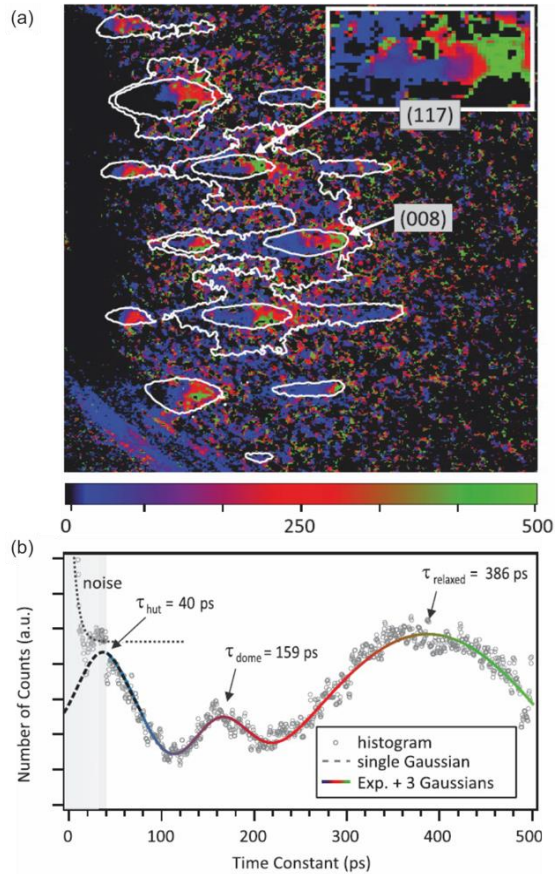

**FIG. 12.** Pixel by pixel analysis of the entire diffraction pattern, the so called lifetime map. (a) Color separation of diffraction spots indicate different cooling time constants for hut, dome, and relaxed clusters. (b) The histogram of the lifetime map analysis clearly shows three distinct cooling time constants, which agree well with the transient spot profile analysis depicted in Fig. 11.

The time constant of all pixels of the diffraction pattern from Fig. 12(a) are sorted in the histogram shown in Fig. 12(b). Three clear maxima can be identified at 40 ps, 160 ps, and 390 ps, which agree well with the recovery time constants for hut-, dome-, and relaxed clusters which were independently determined from the spot profile analysis presented in Fig. 11.

## 3.3 Debye-Waller effect at large momentum transfer: Electron-phonon coupling for Bi(111)

Bismuth is one of the prototypical model systems for studies of laser induced energy transfer from an excited electron system to the lattice system in the time domain. Bi is a semimetal with the conduction band slightly lower in energy than the valence band. The almost vanishing density of states at the Fermi energy results in a low number of free carriers of $10^{17}$ - $10^{19}$ cm$^{-3}$. This makes Bi very sensitive to optical excitations as changes in the electron occupation affects the potential energy surface and trigger atomic motion through displacive excitation. Bismuth exhibits a Peierls distortion which breaks the translational symmetry along the (111) direction with every second Bi atom at a position slightly displaced from the center along the body diagonal of the unit cell. This equilibrium structure, in particular the distance of the two atoms of the basis, can easily be perturbed by electronic excitation.[56,57] resulting in an oscillation of the Bi atoms along the body diagonal, i.e., the coherent symmetric $A_{1g}$ optical phonon mode of the crystal.[58-63]

Depending on the degree of fs-laser optical irradiation vastly different time constants for the excitation process of the Bi lattice were observed. Strong excitation with fluences of more than 6 mJ/cm$^2$ generates so many electron hole pairs that this causes a rapid change of the potential energy surface resulting in non-thermal melting. For fluences of 18 mJ/cm$^2$, the electronic acceleration of the atomic motion occurs as fast as 190 fs, resulting in ultrafast melting, destruction of the Bi-film, and a coherent $A_{1g}$ phonon mode is not observed.[64,65] For fluences lower than 6 mJ/cm$^2$ the lattice response is reversible, the coherent $A_{1g}$ optical phonon mode is excited,[61] and bond softening occurs which results in an inverse Peierls transition.[60,65-67] Subsequently the lattice is heated on slower time scales of 2 - 4 ps [61,64,68-70] through energy transfer from the electron system to the lattice by electron phonon coupling and anharmonic coupling of the $A_{1g}$ mode to acoustic phonons.[71] The vibrational excitation of the surface atoms is even slower: thermal motion of the Bi surface atoms sets in on a timescale of 12 ps and has been attributed to the weak coupling between bulk and surface phonons.[69]

Due to its high atomic mass and weak bonds Bi exhibits a bulk Debye temperature of only $\Theta_D$ = 112 K [72] and thus a large vibrational amplitude upon the thermal motion. These large displacements make Bi an ideal model system to study lattice dynamics upon impulsive optical excitation by means of diffraction techniques. Here we analyze the lattice excitation of the Bi(111) surface through tr-RHEED.

In earlier studies the Debye-Waller effect was employed to follow the onset of atomic motion through the transient intensity changes of the diffraction patterns.[3,5,19,29] Electron diffraction allows for a large momentum transfer due to the possible large scattering angles which result in large intensity changes. Employing all detected diffraction spots of the RHEED pattern for the analysis provides a huge variation of the momentum transfer $\Delta\mathbf{k}$ in diffraction, i.e., a wide range of parallel $k_∥$ and vertical $k_⊥$ momentum transfers are available all at once. Such analysis is presented here.

We followed the excitation of the surface lattice through the Debye-Waller effect $I/I_0 = \exp -\langle(\mathbf{u}\Delta\mathbf{k})^2\rangle$ with the vibrational amplitude $\mathbf{u}$ of the atoms and the momentum transfer $\Delta\mathbf{k}$. For small intensity drop $\Delta I(t) = 1 - I(t)/I_0 < 0.2$ the intensity evolution $I(t)/I_0$ can linearly be converted with an error of less than 6 % in the time constant to a transient change of vibrational amplitude $\mathbf{u}(t)$ applying the linear expansion of the exponential function. This linear expansion, however, becomes inapplicable for intensity drops $\Delta I(t) > 0.2$ which easily occurs for systems with a low Debye temperature, strong excitation, or diffraction at large momentum transfer $\Delta k$. In such cases, the intensity $I(t)$ decays with a time constant which becomes significantly shorter with increasing intensity drop $\Delta I(t)$: a behavior which can easily be misinterpreted as fluence or temperature dependent electron phonon coupling.

Here we used RHEED spots on three different Laue circles, i.e., with different $k_\parallel$ and $k_\perp$, and various laser pump fluences for the excitation of the Bi(111) film in order to analyze the lattice dynamics of the Bi(111) surface. The non-linearity of the exponential function causes the decrease of the time constant $\tau_{int}$ for the decay of RHEED spot intensity from 11 ps to 5 ps with increasing laser fluence $\Phi$ and increasing momentum transfer $\Delta \mathbf{k}$. Irrespective of this large variation of $\tau_{int}$, we obtain a time constant of 12 ps for the heating of the Bismuth surface which is independent of the level of excitation.

The 8 nm thick epitaxial Bi(111) film was grown on a clean Si(111)-(7×7) reconstructed substrate.[6,73] This sample is excited by 800 nm laser pulses at pump powers up to 1200 mW, corresponding to an incidence fluence of $\Phi = 2$ mJ/cm$^2$. A tilted pulse front scheme was used to compensate the velocity mismatch between electron and laser pulse. Here we used electrons of 26 keV with a de Broglie wavelength of $\lambda = 7.6$ pm or momentum $k_0 = 2\pi/\lambda = 82.6$ Å$^{-1}$. They were diffracted at the sample under grazing incidence of 3.4°, i.e., resulting in a vertical momentum transfer of 9.3 Å$^{-1}$ for the specular (00)-spot.

In general, the Debye-Waller effect

$$I(t)/I_0 \exp(-\langle \Delta \mathbf{u}(t)\Delta \mathbf{k}\rangle^2) \qquad (4)$$

with the change of vibrational amplitude $\Delta \mathbf{u}(t) - u_{T0}$ depends on the vector of momentum transfer $\Delta \mathbf{k}_{k,l}$ of the specific diffraction spot $(k,l)$. Only for isotropic displacements of the atoms this expression gives Eq. (4) otherwise it allows the disentanglement of parallel and vertical displacements of the surface atoms.

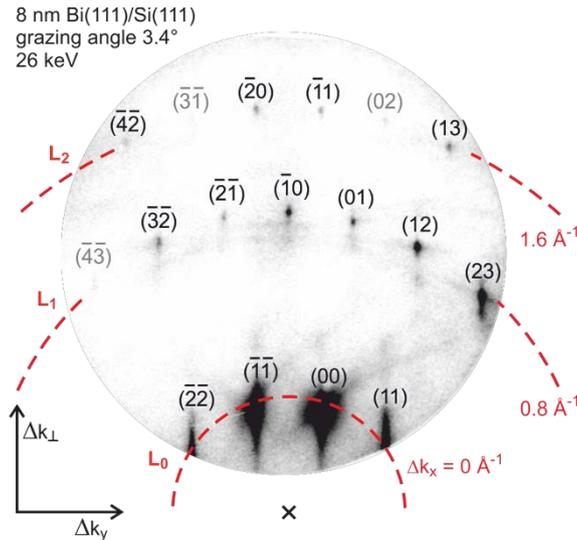

**FIG. 13.** Diffraction pattern of a 8 nm thick Bi(111) film on Si(111) recorded at an electron energy of 26 keV, a grazing angle of incidence of 3.4°, and a sample temperature of $T_0 = 90$K. The vertical $\Delta k_\perp$ and parallel $\Delta k_y$ momentum transfers of the diffracted electrons are indicated. All spots were identified using their Miller indices. The momentum transfer in x-direction along the incident electron beam (see Fig. 4) depends on the order of Laue circle (dashed lines indexed by $L_0$, $L_1$, $L_2$).

Figure 13 shows the diffraction pattern of the Bi(111)-film grown on Si(111) at a sample temperature of 90 K. The momentum transfer $\Delta\mathbf{k}$ is determined for all diffraction spots from diffraction geometry and reciprocal lattice constants. The diffraction pattern is shown in units of $\Delta k_\perp$ (left axis) and $\Delta k_y$ (bottom axis). $\Delta k_x$ increases with the order of Laue circles (dashed lines). Values for $\Delta k_\perp$ cover the range from 7 to 22 Å$^{-1}$. The momentum transfer $|\Delta k_\parallel|$ parallel to the surface is below 8 Å$^{-1}$ for all observed spots. Since $\Delta k_\perp \gg |\Delta k_\parallel|$, our experiment is mainly sensitive to a change of the vibrational amplitude perpendicular to the surface.

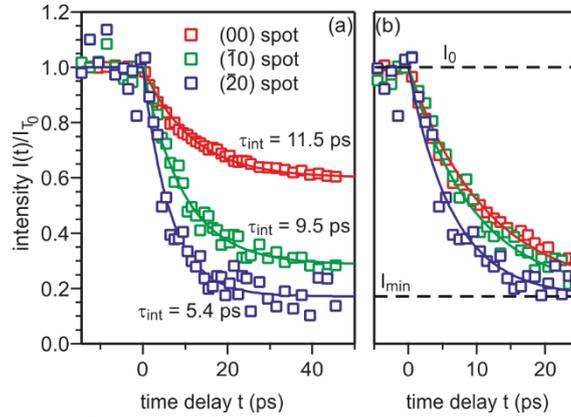

**FIG. 14.** (a) The intensity $I(t)/I_{t0}$ as function of the time delay is shown for three diffraction spots on different Laue circles (red: 0th, green: 1st, blue: 2nd) is shown. The intensity drop $\Delta I_{max}$ increases with momentum transfer from 40 % to more than 80 % while the time constant decreases from 11.5 ps to 5.4 ps. (b) The intensity was normalized to the intensity drop to illustrate the difference in the time constants $\tau_{int}$. The incident pump laser fluence is $\Phi \simeq 1.4$ mJ/cm$^2$.

In Fig. 14(a) the intensity evolution upon excitation with an incident pump laser fluence of $\Phi \simeq 1.4$ mJ/cm$^2$ is shown for diffraction spots on the three Laue circles: the (00)-spot, the ($\bar{1}$0)-spot and the ($\bar{2}$0)-spot. All diffraction spots exhibit an intensity drop that can be described by an exponential function. This intensity drop is caused by the Debye-Waller effect. The intensity decay $I(t)/I_{t0}$ of the three diffraction spots in Fig. 14(a) scales with the squared momentum transfer that increases from 86.5 Å$^{-2}$ for the (00)-spot to 472 Å$^{-2}$ for the ($\bar{2}$0)-spot. The time constant obtained from the exponential fit decreases from 11.5 ps for the (00)-spot to 5.4 ps for the ($\bar{2}$0)-spot. To clearly illustrate this difference of the time constants the normalized intensity drop $\Delta I(t)$ is plotted in Fig. 14(b).

In earlier works (see chapter cooling of film) the transient intensity drop was directly converted into a temperature curve employing a stationary calibration measurement[11,29,30]. Here, we analyze the transient spot intensity without such conversion. For simplicity we apply the Debye model in the high temperature regime ($T \geq \Theta_{D,surf}$) and assume an isotropic MSD $\langle u^2 \rangle$ proportional to the temperature:

$$\langle u^2 \rangle = 3\hbar^2\, T/Mk_B\, \Theta_{D,surf}, \qquad (5)$$

where $\Theta_{D,surf}$ is the effective surface Debye temperature in the framework of individual harmonic oscillators ($\Theta_{D,surf}$ = 47 K for the Bi(111) surface[29,74,75]) and $M$ the atomic mass of Bi. We also assume an exponential increase of MSD, i.e., an exponential rise of temperature $T(t)$ to a maximum temperature $T_0 + \Delta T_{max}$, with a time constant $\tau_T$ :

$$T(t) = T_0 + \Delta T_{max} \cdot \Theta(t) \, (1 - \exp(-t/\tau_T)). \qquad (6)$$

The intensity is:

$$I(t)/I_{t0} = \exp\left[-\alpha \Delta T_{max} \, \Theta(t) \, (1 - \exp(-t/\tau_T))\right] \qquad (7)$$

with $\alpha = \hbar^2 \Delta k^2 / M k_B \Theta^2_{D,surf}$. For small values of $\alpha \Delta T_{max}$ we can safely use a linear approximation of the exponential because the higher order terms in the expansion are negligible small:

$$I(t)/I_{t0} \simeq 1 - \alpha \Delta T_{max} \, \Theta(t) \, (1 - \exp(-t/\tau_T)). \qquad (8)$$

With this approximation, the maximum intensity drop is $\Delta I_{max} = \alpha \Delta T_{max}$ and the time constant $\tau_{int}$ - as experimentally determined from the transient intensity decay - is almost the same as that $\tau_T$ from the temperature curve. The question arises up to what arguments $\alpha \Delta T_{max}$ we can use this linear approximation?

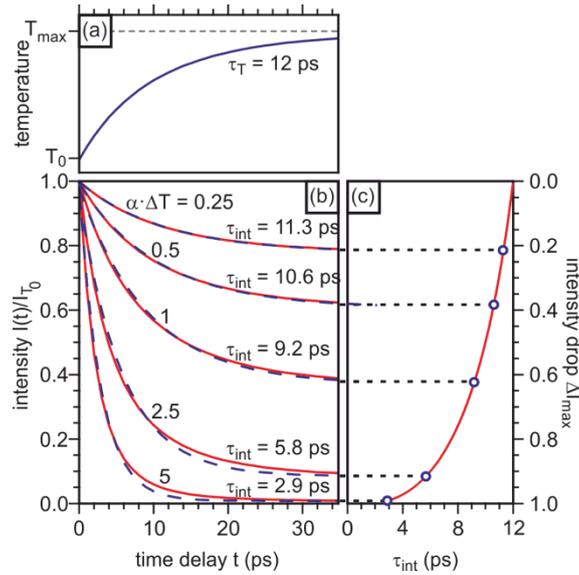

**FIG. 15.** The intensity for an exponential temperature rise by $\Delta T_{max}$ with time constant $\tau_T = 12$ ps was modeled and is plotted as function of the time delay with different values of $\alpha \Delta T_{max}$ (solid lines) on the left side. The curves are fitted by an exponential decay function (dashed lines). On the right side, the time constant obtained from the fit is plotted over the intensity drop $\Delta I_{max}$. With increasing values for $\alpha \Delta T_{max}$ the intensity drop becomes larger and the fitted time constants decreases dramatically from 12 ps for $\alpha \Delta T_{max} \approx 0$ to 2.9 ps for $\alpha \Delta T_{max} = 5$.

We modeled the intensity to obtain the time constant $\tau_{int}$ in dependence of the intensity drop $\Delta I_{max}$. An exponential temperature rise with a time constant of $\tau_T = 12$ ps (see Fig. 15(a) and observed in ref. 69) is converted into the corresponding intensity $I(t)$ using Eq. (7). $I(t)$ is plotted in Fig. 15(b) as function of the time delay for 5 different values of $\alpha \Delta T_{max}$ (solid lines) and fitted with an exponential decay function as given by Eq. (3) (dashed lines). For small values $\alpha \Delta T_{max} = 0.2$ the calculated intensity $I(t)$ exhibits almost the same behavior like $T(t)$ and is well described by the fit function (Eq. (3)). The intensity drop $\Delta I_{max}$ is ≤ 18% and the time constant obtained from the exponential fit (dashed line) $\tau_{int}$ = 11.3 ps deviates only by 6% from $\tau_T$. With increasing values for $\alpha\Delta T_{max}$, however, the time constant obtained from the exponential fit $\tau_{int}$ (dashed lines in Fig. 15.) decreases. In the right panel of Fig. 15 the fitted time constant $\tau_{int}$ is plotted as function of the intensity drop $\Delta I_{max}$. For $\Delta I_{max}$ approaching unity, i.e., drop to intensity to almost zero, the time constant $\tau_{int}$ decreases to 3 ps and less. We therefore have to expect strongly varying experimental time constants $\tau_{int}$ depending on the degree of excitation ($\Delta T_{max}$) or momentum transfer ($\alpha$). The varying time constants of 5.4 – 11.5 ps obtained for the different orders of Laue circles shown in Figure 14 are thus explained by the correlation of $\Delta I_{max}$ and $\tau_{int}$ as shown in Fig. 15(c). The correct time constant of the temperature rise $\tau_{int}$ can only be found by extrapolation to $\Delta I_{max}$ = 0. Therefore, under our diffraction conditions at large momentum transfer $\Delta k$ and large intensity drop $\Delta I_{max}$ the time constants $\tau_{int}$ can be much shorter than $\tau_T$. In the following we perform a thorough Debye-Waller analysis in order to prove, that the pre-conditions for such analysis are still valid.

$$I(t)/I_{t0} \simeq 1 - \alpha\Delta T_{max} \Theta(t) (1 - \exp(-t/\tau_T)). \tag{8}$$

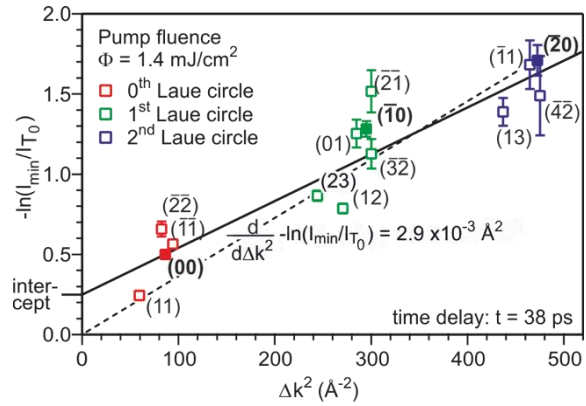

**FIG. 16.** The negative logarithm of the minimum intensity $I(T_{max})/I_{T0}$ is plotted as function of $\Delta k^2$ for all diffraction spots at a time delay of t = 38 ps. Data from the different Laue circles are plotted in different colors. If applying kinematic scattering theory a linear fit through the origin is expected (dashed line). The solid line gives a better fit to the data and the intercept is explained by multiple scattering effects following literature [75,76,77].

From the change of spot intensity, we obtain information about the change of the MSD:

$$-\ln(I(T)/I_{T0}) = 1/3\, \Delta k^2\, (\langle u^2(T)\rangle - \langle u^2_{T0}\rangle). \tag{9}$$

From kinematic diffraction theory,[24,25] we expect a linear dependence of the negative logarithm of the intensity $-\ln(I(T)/I_{T0})$ as function of $\Delta k^2$ with a y-axis intercept equal to zero Eq. (9). The slope $-d(\ln(I(T)/I_{T0}))/d(\Delta k_2)$ is equal to one third of the change of the MSD $\Delta\langle u^2\rangle = \langle u^2(T_{max})\rangle - \langle u_{T0}^2\rangle$ or, if the effective surface Debye temperature is known (here $\Theta_{D,surf}$ = 47 K), proportional to the temperature rise $\Delta T_{max}$, respectively. Figure 16 depicts -$\ln(I_{min}/I_{T0})$ for all diffraction spots plotted as function of the squared momentum transfer $\Delta k^2$.

The value $I_{min}$ is the minimum intensity obtained from the fit for the maximum transient temperature. The expected behavior for kinematic diffraction theory and isotropic vibrational motion is plotted as dashed line. The data are, however, better described by a linear fit with a y-axis intercept > 0. Such positive intercept was also observed in transmission electron diffraction experiments[75-77] and is caused by multiple scattering effects. The offset observed in transmission electron diffraction was found to be proportional to the temperature change as well and is explained by dynamical two beam diffraction theory.

Though the data in Fig. 16 scatter around the linear slope, we did not find any systematic deviations as function of parallel $\Delta k_\parallel$ or vertical $\Delta k_\perp$ momentum transfer. This justifies the pre-assumption of an isotropic thermal motion. Thus, the present data do not provide insight into any potential non-equipartition in parallel or vertical vibrational amplitude. Finally, we obtain a change of the MSD at $t = 38$ ps that is $\Delta \langle u^2 \rangle = 8.8 \cdot 10^{-3}$ Å$^2$.

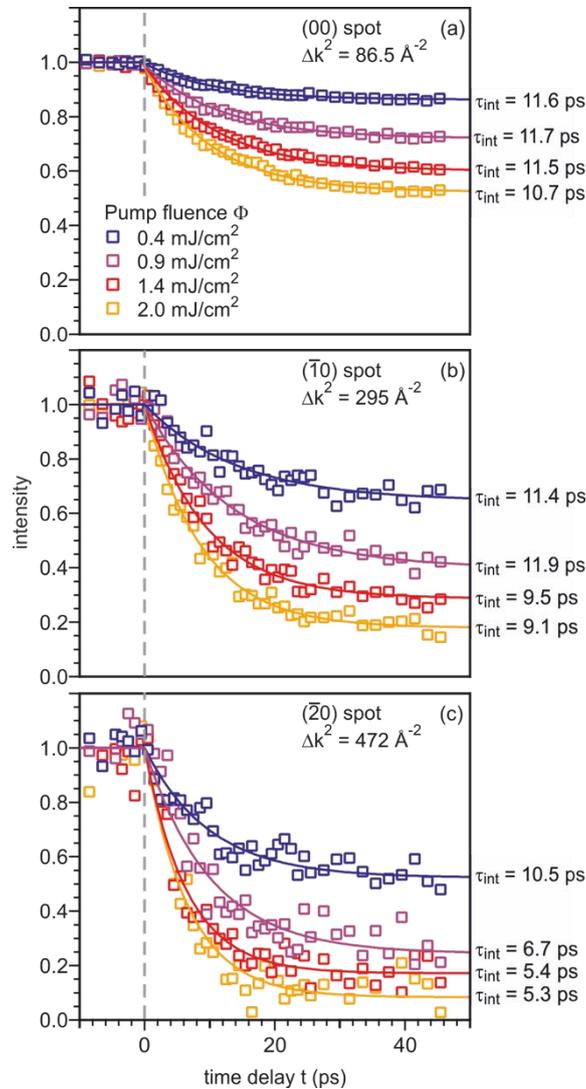

**FIG. 17.** Transient intensity drop of the (00), ($\bar{1}$0), and ($\bar{2}$0) spot shown in (a), (b), and (c) for increasing incident pump fluence, respectively. The time constant for an exponential fit to the data decreases with increasing momentum transfer $\Delta k$ and increasing excitation density.

The intensity drop $\Delta I_{max}$ depends on the absorbed energy that was changed by varying the pump fluence. In Fig. 17 the intensity as function of the time delay is plotted for three diffraction spots (same as in Fig. 14) and four different pump fluences $\Phi$ between 0.4 and 2 mJ/cm$^2$. The intensity drop $\Delta I_{max}$ becomes larger with increasing pump fluence for all diffraction spots.

For each pump fluence a Debye-Waller analysis like in Fig. 16 was performed. The change of the MSD $\Delta\langle u^2 \rangle$ rises linear with the pump fluence [78]. From this we conclude that the absorbed energy is proportional to the pump fluence and the vibrational motion of the atoms is still in the harmonic regime of the potential. For the maximum laser pump fluence of $\Phi = 2$ mJ/cm$^2$ the MSD increases by $\Delta\langle u \rangle = 11.9 \times 10^{-3}$ Å$^2$. This corresponds to a asymptotic temperature rise of $\Delta T_{max} = 72$ K.

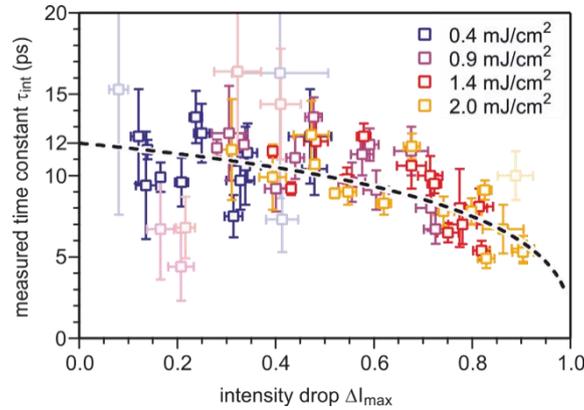

**FIG. 18.** The measured time constant $\tau_{int}$ is plotted versus the intensity drop $\Delta I$ for all diffraction spots and all pump fluences. The light symbols represent weak diffraction spots with strong noise and large error bars. The dashed line is the expected behavior for a temperature rise $\Delta T$ with a time constant of 12 ps.

Increasing pump fluence and analysis of spots with larger momentum transfer $\Delta k$ result in strongly enhanced intensity drops $\Delta I_{max}$ and thus much shorter time constants $\tau_{int}$. The modeling shown in Fig. 15 explains this correlation well. Fig. 18 summarizes all the experimental results and compares them with the expected behavior of $\tau_{int}(\Delta I)$ shown in Fig. 15 (dashed line). For each diffraction spot the time constant determined from the fit is plotted over the intensity drop for all four pump fluences. For the determination of the time constant $\tau_T$ of the temperature rise we modeled $\tau_{int}(\Delta I_{max})$-curves for different values of $\tau_T$ and found a minimum standard deviation for $\tau_T = (12.0 \pm 0.4)$ ps. No evidence for a dependence of excitation time constant $\tau_T$ on the excitation level is found in the regime of weak excitation with incident pump fluences $\Phi \leq 2$ mJ/cm$^2$.

We thus observe a time constant for heating of the surface atoms which is more than 4 times larger than values reported for the bulk under conditions comparable to our incident laser fluences.[69] Thus, the surface is not following the excitation of the bulk. Instead the thermal excitation of the surface atoms occurs delayed on a time scale of 12 ps which we attribute to a reduced electron phonon coupling at the surface. The Bi(111) surface exhibits a pronounced electronic surface state.[79-84] This surface state is easily populated upon fs IR irradiation.[85] The population of excited electrons in this surface state exhibit a lifetime comparable to the thermalization time constant observed in our experiment (see Fig. 5 (a) of ref. [85]). We conclude that slow energy transfer from the electronic surface state to the surface atoms is the dominant mechanism for the thermal excitation of the surface.

### 3.4 Damping of Vibrational Excitations of Monolayer Adsorbate Systems

Through excitation of localized vibrational modes in 2D adsorbate layers, it is possible to feed energy into a solid-state system at very high spatial selectivity. Transfer and dissipation of the deposited vibrational energy are topics of general interest, both from a fundamental and applied viewpoint, e.g., for controlling heat transfer through interfaces[86] or chemical reactions[87] at surfaces. Usually the lifetime of vibrational modes is studied by means of infrared,[88] sum frequency,[89-91] or Raman spectroscopy.[92] While each of these techniques has its specific advantages, the conservation laws and selection rules limit each technique to specific modes and regions of reciprocal space. Moreover, for heavy adsorbates, the vibrational frequencies are in the far infrared and thus difficult to access experimentally. In addition, diffraction methods have the advantage of being able to access spatial information as well as low frequency vibrations since the vibrational amplitude $u$ is proportional to $1/\omega$. Thus, topics like mode coupling among the adsorbate layer and towards the substrate which ultimately are responsible for the relaxation of the vibrational excitation can experimentally be accessed.

These processes have been studied for the vibrational dynamics of an ordered atomic layer of Pb atoms adsorbed on Si(111). Due to the large atomic mass of Pb and its low bulk modulus, very soft vibrational modes are expected in the monolayer Pb film. We will show that these modes couple to the acoustic phonons of the Si substrate only in a small sector of the two-dimensional Brillouin zone. Therefore, the acoustic phonons possess an unusually long lifetime of several nanoseconds, and thus lend themselves as ideal objects for the study of relaxation by intermode coupling.

We employ the vectoral expression of the Debye-Waller effect:

$$I(\mathbf{k}) = \exp{-\langle(\mathbf{k}\mathbf{u})^2\rangle} = \exp{-(\langle(\mathbf{k}_\parallel \mathbf{u}_\parallel)^2\rangle + \langle(\mathbf{k}_\perp \mathbf{u}_\perp)^2\rangle)}, \qquad (10)$$

where the scalar product of the vibrational displacement amplitude $\mathbf{u}$ and the momentum transfer $\mathbf{k}$ determines the sensitivity to a specific eigenmode. Because the (00) spot has no parallel momentum transfer $\mathbf{k}_\parallel = 0$ it is insensitive for any parallel component $\mathbf{u}_\parallel$ of the vibrational modes and only sensitive to the vertical component $\mathbf{u}_\perp$. Higher-order spots possess a parallel component of the momentum transfer $\mathbf{k}_\parallel \neq 0$ and are, therefore, also sensitive to the parallel component $\mathbf{u}_\parallel$ of the vibrational modes.

Here, we disentangle contributions from high-frequency in-plane $\mathbf{u}_\parallel$ from low-frequency out-of-plane $\mathbf{u}_\perp$ polarized vibrational modes which exhibit clearly different dynamics in the time domain. Such approach allowed us to follow the conversion of the initial electronic excitation via strong electron-lattice coupling, the conversion of vibrational modes, and the vibrational energy dissipation into the substrate.

The Pb layer with a coverage of 4/3 monolayer is prepared in the (√3×√3) phase by deposition on a clean Si(111)-(7×7) substrate at 90 K and subsequent annealing up to 700 K. [93-95] Laser pulses with an incident fluence of 4.0 mJ/cm$^2$ at a wavelength of 800 nm (1.55 eV) are used as pump pulses. Optical excitations of the Si substrate with its direct band gap of 3.4 eV are still negligible at that fluence. The tr-RHEED experiments were performed with 7 keV electrons. The velocity mismatch reduces the temporal resolution to only 40 ps [55,96].

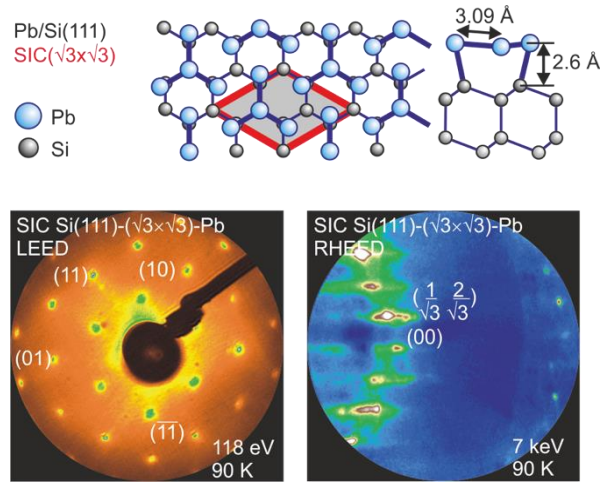

**FIG. 19. (a):** Schematics of the atomic geometry of the SIC Si(111)-(√3×√3)-Pb reconstruction [93-95] **(b)** LEED and **(c)** RHEED pattern of Si(111)-(√3×√3)-Pb reconstruction on Si(111). The RHEED pattern shows integer order and fractional order spots located on the zero order Laue circle.

The relaxed geometric structure of the Pb (√3 × √3) phase on Si(111) is presented in Fig. 19(a). The Pb layer shows an adsorption height of 2.6 Å [96]. Each (√3×√3) unit cell possesses four Pb atoms, i.e., a saturation coverage of 4/3 of a Si(111) monolayer (ML = 7.8 × 10$^{14}$ atoms/cm$^2$). Three Pb atoms are bonded in a T1 site to the dangling bonds of the Si(111) surface, the remaining Pb atom in a T4 site bonds solely to the three other Pb atoms and is located in their center as sketched in Fig. 19(a). The two-dimensional electronic states induced by the metallic Pb layer are partially occupied and electrons from the filled Pb bands or from the Si substrate can be excited into the unoccupied Pb bands using infrared photons [96]. As the Si band gap prevents their diffusion into the substrate, the electrons will be deexcited by electron-electron and electron-phonon scattering within the Pb monolayer, thus exciting Pb phonons. LEED and RHEED patterns are shown in Figs. 19(b) and 19(c), respectively.

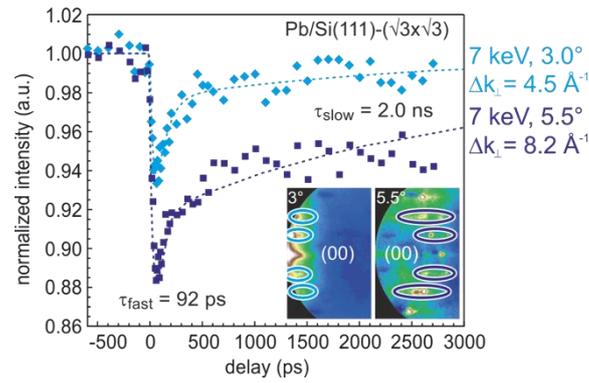

**FIG. 20.** The transient intensity drop of the (√3×√3) spots upon excitation by a femtosecond-laser pulse exhibits a biexponential recovery with a fast ($\tau_{fast} \cong 100$ ps) and slow component ($\tau_{slow} \cong 2$ ns). Data for grazing angles of 3° and 5.5° are shown. Inset: RHEED patterns of the Pb-induced (√3×√3) phase on Si(111) at 90 K. Those (√3×√3) spots used for the analysis are encircled.

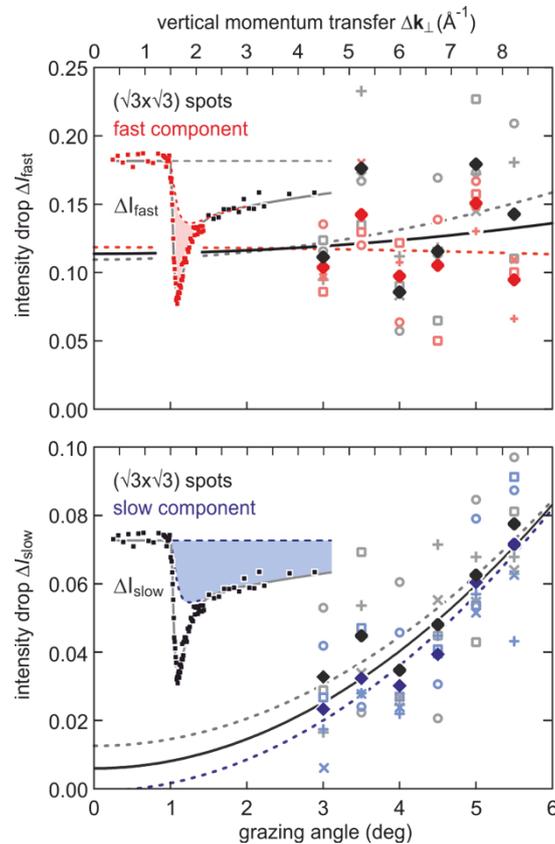

**FIG. 21.** Coefficients for the fast $\Delta I_{fast}(\theta)$ and slow $\Delta I_{slow}(\theta)$ component of the Debye-Waller effect as a function of the grazing angle θ of the electron beam. Open symbols display intensity drops of individual (√3×√3) diffraction spots. Solid symbols display averaged data. Solid lines are a parabolic fit to the data. Dotted lines give error margins of fit.

In tr-RHEED the transient intensities of individual diffraction spots were analyzed by integration of a small elongated region of interest around each spot, which are marked in Fig. 20 and normalized by the number of pixels. The (00) spot was excluded from the analysis as it may be affected by contributions from the Si substrate and still existing Pb islands. Multiple runs were performed at varying grazing angles between 3° and 5.5°, thus allowing for a variation of perpendicular momentum transfer from $\Delta k_\perp = 4.5$ to 8.2 Å$^{-1}$. Figure 20 shows for grazing angles of incidence of 3° and 5.5° transients of the intensity of a ($\sqrt{3} \times \sqrt{3}$) diffraction spot, displaying a sharp drop, followed by a slow recovery that extends over nanoseconds.

The bi-exponential recovery of the ground state is described by a fast $\Delta I_{fast}$ and slow $\Delta I_{slow}$ component of the intensity drop:

$$I(t)/I_0 = 1 - \Theta(t) [\Delta I_{fast} \exp(-t/\tau_{fast}) + \Delta I_{slow} \exp(-t/\tau_{slow})], \qquad (11)$$

where $\Theta(t)$ is the Heaviside function that defines temporal overlap at $t = 0$. The fast contribution exhibits a decay constant $\tau_{fast} \leq 100$ ps, while the much slower contribution relaxes with a decay constant $\tau_{slow} \geq 2000$ ps. Fitting the experimental data with Eq. (11) yields very similar time constants for the recovery of intensity for all diffraction angles, and hence, by averaging all angles, we find the values of $\tau_{fast} = 92 \pm 8$ ps and $\tau_{slow} = 2007 \pm 267$ ps. This fit is shown in Fig. 20 as dashed lines together with the data for incident angles of $\theta = 3.0°$ and $\theta = 5.5$. We note that the time constants $\tau_{fast}$ and $\tau_{slow}$ are markedly different, i.e., the corresponding relaxation processes must be significantly different.

The corresponding intensity coefficients $\Delta I_{fast}(\theta)$ and $\Delta I_{slow}(\theta)$ are shown in Fig. 21 as a function of the grazing angle $\theta$. We notice that $\Delta I_{fast}(\theta)$ is independent (within statistical error) of $\theta$, whereas $\Delta I_{slow}(\theta)$ increases with $\theta$. According to Eq. (10), the intensity coefficient is proportional to the mean square of the Pb displacement vectors projected onto the momentum transfer $\Delta \mathbf{k}$. Because $\Delta \mathbf{k}$ is predominantly perpendicular to the surface in our experiment, we are mostly sensitive to perpendicular adsorbate vibrations. From Fig. 21(a) we are led to the conclusion that the Pb displacements, contributing to $\Delta I_{fast}$, are almost entirely within the surface plane, and hence insensitive to the amount of perpendicular momentum transferred in the diffraction. We determine a dominant change of the parallel mean squared displacement of $\Delta \langle u_{\parallel}^2 \rangle = 0.025$ Å$^2$.

The long-lived vibrational excitations in the Pb layer that contribute to $\Delta I_{slow}$ therefore has to exhibit a strong perpendicular component in order to explain the observed angular dependence with $\Delta I_{slow} \sim \Delta k_{\perp}^2$. The parabolic dependence on vertical momentum transfer $\Delta k_{\perp}$ is expected from the expansion of Eq. 5 for small intensity decreases $\Delta I < 0.2$. We determine a change of the vertical mean squared displacement of $\Delta \langle u_{\perp}^2 \rangle = 0.0016$ Å$^2$. From the almost vanishing intercept of $\Delta I_{slow}(k_{\perp})$ with the ordinate at $k_{\perp} = 0$ we conclude a small parallel component for the slow contribution only. A parallel component of the long-lived vibrations, however, cannot be completely ruled out, as the experimental diffraction geometry with $k_{\perp} > k_{\parallel}$ results in a higher sensitivity to perpendicular as to parallel displacements of the Pb atoms.

Because the data in Fig. 21 suggest that initially no out-of-plane modes are excited, the decay of the optical Pb modes into substrate modes must be accompanied by a conversion of a small fraction of the excitation into low-energy out-of-plane modes. From the value of $\Delta I_{slow}$ together with a measurement of the static Debye-Waller effect, we estimate the rise of temperature in the Pb system of $\Delta T = 20$ K after thermalization of the phonon system for times beyond 100 ps.

In summary, we find that upon irradiation with the fs-laser pulse, initially only modes with dominant parallel vibrational amplitudes $u_\parallel \gg u_\perp$ are excited. These modes convert on a timescale of $\tau_{fast} \leq 100$ ps to modes that exhibit a dominant vertical vibrational amplitude $u_\perp \gg u_\parallel$.

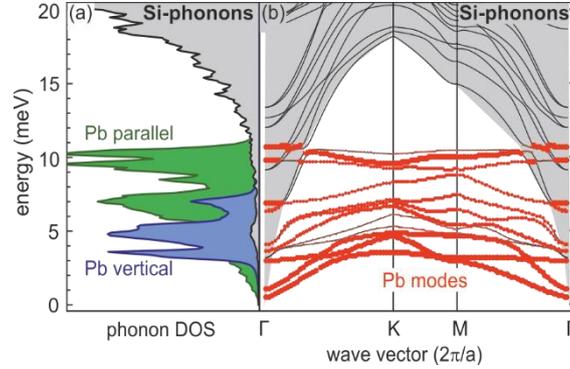

**FIG. 22.** (a) Mode specific phonon DOS for the Pb layer and for the Si substrate. Parallel and vertical modes are shown in green and blue, respectively. (b) Phonon dispersion for the Pb-layer (red dots) and the Si substrate in gray. Figure adopted from [96]

The subsequent excitation of the fast and slow modes and their two significantly different de-excitation time constants can be explained through electron-phonon coupling in the Pb layer and phonon-phonon coupling from Pb layer to the Si substrate. Figure 22(a) depicts the phonon dispersion of the Pb layer in solid red lines as function of wave vector with a maximum energy of the phonons at 10 meV. Most of these modes may be denoted as optical modes with a finite energy of 4 to 10 meV at the Γ-point. There are only three acoustic phonon states with $E(\Gamma) = 0$ at the zone center and $E(K) = 5$ meV at the zone boundary.

Only a small fraction of the phonon phase space can be excited through the excited electron system since the conservation of energy and momentum must be obeyed. Typical values for the dispersion in the electron system of the states in the Pb layer are $\Delta E_{el}/\Delta q \sim 0.1 - 2$ eV/Å$^{-1}$ (see Fig. 3 in ref[96]) while typical values for the generation of phonon are smaller than $E_{ph}/q < 0.005$ eV/Å$^{-1}$ as shown in Fig. 22(a). As a consequence, only optical phonons with almost vanishing wave vector and finite energy fulfill the requirement of conservation of energy and momentum with q < $E_{ph}/0.1$ eV/Å$^{-1}$ ~ 0.05 Å$^{-1}$ during excitation. The acoustic phonons already carry a too large momentum at finite energy and can thus not be excited.

The mode selected density of states (DOS) are shown in Fig. 22(b) and reveals, that the initially excited high frequency optical modes exhibit displacements only parallel to the surface plane without vertical component. These modes are observed as fast component in diffraction. The modes with vertical displacements can be associated with the zone boundary optical and acoustic branches which initially are not excited. These modes become populated through mode conversion.

In the framework of the diffuse mismatch model the cooling towards the Si substrate is determined by the overlap of the Pb and Si phonon DOS as shown in Fig. 22(b). As the parallel modes $\mathbf{u}_\parallel$ exhibit large overlap with the Si modes they decay quickly on a 100 ps time scale into the Si substrate. In contrast, the overlap of the vertical modes $\mathbf{u}_\perp$ with the Si modes is much smaller and consequently the decay of these modes needs more than 2 ns which has also been corroborated by molecular dynamics simulations [96].

## 3.5 Driven phase transitions: Melting of a CDW in atomic wire system Si(111)-In (8x2)↔(4x1)

Due to its unique and peculiar properties, the indium atomic wire system is ideally suited for the study of structural and electronic dynamics at surfaces. This system exhibits an inherent Peierls instability manifesting itself in a 1$^{st}$ order phase transition between an insulating (8×2) ground state and metallic (4×1) high temperature state. This structural transition can non-thermally be driven through an optical excitation and subsequently is trapped for nanoseconds in a supercooled metastable state. The indium atomic wire system is prepared by self-assembly under ultra-high vacuum conditions [97-103]. In situ deposition of a monolayer (1 ML is equivalent to $7.83 \times 10^{14}$ cm$^2$) of indium atoms on Si(111) substrates at a sample temperature of 700 to 750 K creates the (4×1) In/Si(111) reconstruction. The metallic high temperature phase of this atomic wire system is composed of two parallel zig-zag chains of indium atoms with a (4×1) unit cell [98,104] as is sketched in Fig. 23(a). The corresponding low energy electron diffraction (LEED) pattern is shown in the right panel of Fig. 23(a) with its 3-fold symmetry arising from three rotational domains at the hexagonal (111) surface. Upon cooling the system undergoes a reversible transition from the metallic high temperature state to the insulating ground state [99,105,106] at $T_c = 130$ K [102,107,108] which is accompanied with the opening of a band gap of $E_{gap} = 0.2$ eV [102,109] and the formation of a charge density wave (CDW). In the ground state the zig-zag chains of indium atoms are broken and they rearrange into distorted hexagons [98,104], as sketched in Fig. 23(b). Upon this phase transition the maximum change of geometric position of the In atoms in the surface unit cell is less than 0.1 Å only [110]. This Peierls-like transition is characterized by symmetry breaking in both directions which is facilitated through soft, shear- and rotational-phonon modes, with frequencies of $\nu_{shear} = 0.54$ THz and $\nu_{rot} = 0.81$ THz, respectively [104,111-114]. The surface periodicity doubles along and normal to the wires and the size of the unit cell increases to (8×2). This change becomes obvious in the LEED pattern in the right panel of Fig. 23(b) through the appearance of additional spots at 8-fold position between the 4-fold spots. The appearance of 2-fold streaks emerge from the broken correlation of the 2-fold periodicity in neighbored wires. The anisotropic nature of the indium atomic wire system becomes immediately apparent in scanning tunneling microscopy (STM). Figure 24(a) displays a filled-state STM image from ref [102] from the indium wire surface at $T = 135$ K, i.e., at $T_c$. Extended and parallel wires both with (8×2) and (4×1) reconstruction are present. Employing scanning tunneling spectroscopy both at the (8×2) and (4×1) structure - as shown in Figs. 24(b) and 24(c) - reveals the opening of a bandgap of $E_{gap} = 0.16$ eV for the low temperature (8×2) structure - see Fig. 24(d) - and is indicative for the formation of a charge density wave and the metal to insulator transition [102].

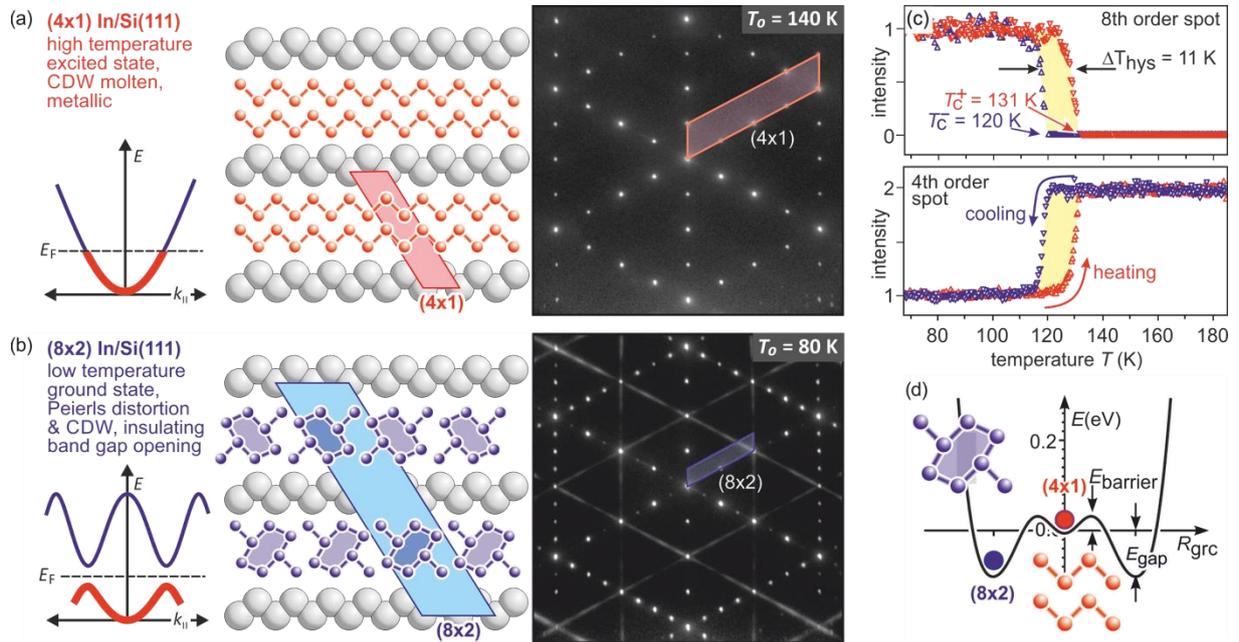

**FIG 23.** (a) The metallic high temperature (4×1) state is composed of In atoms arranged in double zig-zag chains. The LEED pattern depicts the (4×1) reconstruction in three rotational domains. (b) The insulating (8×2) ground state exhibits a Peierls distortion with the formation of a CDW and opening of a band gap. The In atoms are rearranged in distorted hexagons. The (8×2) LEED pattern clearly shows the periodicity doubling along and perpendicular to the wires. The 8-fold spots and 2-fold streaks are indicative for the ground state structure. (c) RHEED intensity of the (8×2) spot (upper panel) and the (4×1) spot (lower panel) as function of temperature. Upon heating the intensity of the (8×2) spot drop to the background at $T_c$. Cooling with the same rate leads to the transition back into the (8×2) reconstruction. The intensity of the (4×1) spot rises upon heating at $T_c$, reflecting the change of atom position in the unit cell. The temperature cycling of both spots exhibits a hysteresis of 11 K. (d) Potential energy surface obtained through density functional calculations as function of a generalized reaction coordinate Q describing the transition from the (4×1) and the (8×2) phase. The blue and red dot indicate the (8×2) ground state and the metastable (4×1) state, respectively.

The equilibrium phase transition has been followed during quasi stationary rise of temperature from 70 K to 180 K where the sharp drop of intensity of the 8-fold spots to zero (see upper panel of Fig. 23(c) is indicative for the transition from the (8×2) ground state to the (4×1) high temperature state. At the same time the intensity of the 4-fold spots sharply rises by a factor of two and thus indicating the structural transition (see lower panel of Fig. 23(c). During slow temperature cycling a hysteresis of the high temperature (4×1) and low temperature (8×2) states is observed upon heating and cooling as shown in Fig. 23(c). The width of the hysteresis is independent on the cooling/heating rate d$T$/d$t$ [107]. Such behavior is evidence of a first-order phase transition, i.e., a non-continuous transition with both states separated by a small energy barrier.

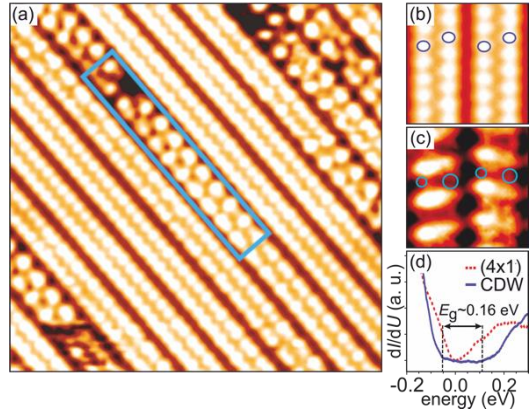

**FIG. 24.** (a) STM micrograph under constant current conditions taken at 100 K. Both (8×2) and (4×1) reconstructed indium wires can be seen. (b) The metallic (4×1) wires are composed of In atoms arranged in two parallel zig-zag chains. (c) Instead, in the insulating (8×2) wires the chains are broken up and distorted hexagons of In atoms form. (d) STS spectra for the (8×2) and (4×1) reconstructed wires. While the (4×1) exhibit metallic behavior (dashed red line), the (8×2) clearly shows opening of a bandgap of $E_{gap} = 0.16$ eV (solid blue line). Data courtesy of H.W. Yeom and with permission from ref. [102].

### 3.5.1 Photo induced phase transition

This 1$^{st}$ order phase transition can be triggered by impulsive optical excitation through intense laser pulses (1 to 10 mJ/cm$^2$ on a femtosecond time scale of 20 to 200 fs). The sudden and massive optical excitation of the electron system transiently changes the potential energy surface for the atoms position in the lattice. This provokes accelerating forces on the atoms ultimately causing the structural transition. This photo-induced transition is demonstrated in Fig. 25 where panel (a) depicts the RHEED pattern of the (8×2) ground state prior to optical excitation at negative pump-probe delays $\Delta t < 0$ at a temperature $T_0 = 30$ K, i.e., well below $T_c = 130$ K. The pattern taken at $\Delta t = 6$ ps, i.e., after optical excitation through a fs-laser pulse with a fluence of $\Phi = 6.7$ mJ/cm$^2$, is shown in panel (c) and exhibits clear differences. The transient changes of spot intensity become more obvious in the difference pattern in panel (b) depicting intensity gains (red) and losses (blue) in a false color representation. All 8-fold spots and 2-fold streaks (indicative for the ground state) disappeared while 4-fold spots (indicative for the high temperature state) gained intensity. The complete transition from the (8×2) ground state to the (4×1) excited state is also reflected by the clear changes in the two representative spot profiles shown for a (8×2) and (4×1) spot in blue and red, respectively.

### 3.5.2 Super-cooled excited state

Surprisingly the excited (4×1) state is stable for ns and only slowly recovers the (8×2) ground state as shown in Fig. 28(f), where the intensity of a (4×1) spot is plotted for long pumpprobe delays $\Delta t$. As we show later, the indium surface layer cools via heat transport on a $\tau_{cool} = 30$ ps timescale to the substrate temperature of $T_0 = 30$ K. We thus can safely exclude a slow thermal recovery of the (8×2) ground state. This long lived (4×1) state is explained through the nature of this phase transition: in general, a first-order transition exhibits a barrier between the two states hindering the immediate recovery of the ground state. This picture is corroborated through density functional calculations of the potential energy surface (PES). Figure 23(d) depicts this PES as function of a generalized reaction coordinate Q obtained by superimposing the soft shear and rotary phonon eigenvectors that transform between the (4×1) and the (8×2) phase [98,104]. We found the transition from the (4×1) phase to the (8×2) structure to be hampered by an energy barrier of $E_{barrier}$ = 40 meV (see Fig. 23). At temperatures below $T_c$ this barrier hinders the immediate recovery to the (8×2) ground state: A longlived metastable and supercooled excited phase is stabilized and trapped in a state far from equilibrium for few nanoseconds [115]. In analogy to a supercooled liquid, one might even expect the freezing, i.e., the transition back to the (8×2) ground state, to be facilitated by condensation nuclei, possibly in form of adsorbates. To verify this assumption experimentally, we monitored the phase transition dynamics upon controlled adsorption of molecules from the residual gas. The transient intensity evolution of the (8×2) (black to green dots) and (4×1) spots (red to yellow dots) is plotted in Fig. 26(a) for various adsorption times $t_{ad}$. With increasing adsorbate coverage, we observed a strong decrease in the time constant, as depicted in Fig. 26(b).

The shortest observed time constant was $\tau$ = 54 ps for an adsorption time of $t_{ad}$ = 75 min. The solid line shows a fit to a $1/t_{ad}$ behavior. Obviously, the adsorption from the residual gas drastically shortens the recovery time of the (8×2) ground state by almost a factor of 10. Sticking to the analogy with a supercooled liquid, the insertion of seeds, i.e., condensation nuclei, initiates the freezing, which then propagates with constant velocity. Here, freezing means recovery of the (8×2) ground state. Because of the highly anisotropic nature of the indium-induced Si surface reconstruction, this phase front propagates only onedimensionally along the direction of the indium chains. Therefore, the velocity of the phase front $v_{(8×2)}$ within the onedimensional In wire and the averaged distance $l_{ad}$ between the condensation nuclei determine the time constant t for the complete recovery of the (8×2) ground state: $\tau = l_{ad} / (2\ v_{(8×2)})$; as sketched in Fig. 27. In addition, assuming a linear relation between adsorbate coverage $\Theta_{ad}$ and the time $t_{ad}$, the distance between the adsorbates in one row obeys $l_{ad} \propto t_{ad}^{-1}$; consequently, it holds $\tau \propto t_{ad}^{-1}$. This is indeed the experimental finding shown in Fig. 26(b). An estimate for the distance lad between adsorbates in one individual row can be obtained from the shift of critical temperature $T_c$ as a function of the adsorbate density $\Theta_{ad}$. We observed $\Delta T$ = + 40K after adsorption for $t_{ad}$ = 75 min. According to Lee and Shibasaki, such a change in $T_c$ is induced by an adsorbate density of $\Theta_{ad} = 6×10^{12}$ cm$^{-2}$ as determined by scanning tunneling microscopy [108,116]. The distance lad between the adsorbates, together with the measured time constant t, are sufficient to determine the lower limit of the phase front velocity $v_{(8×2)}$. The present experimental data result in a value of $v_{(8×2)}$ = 82 m/s [109].

Invited Review for Structural Dynamics 2023

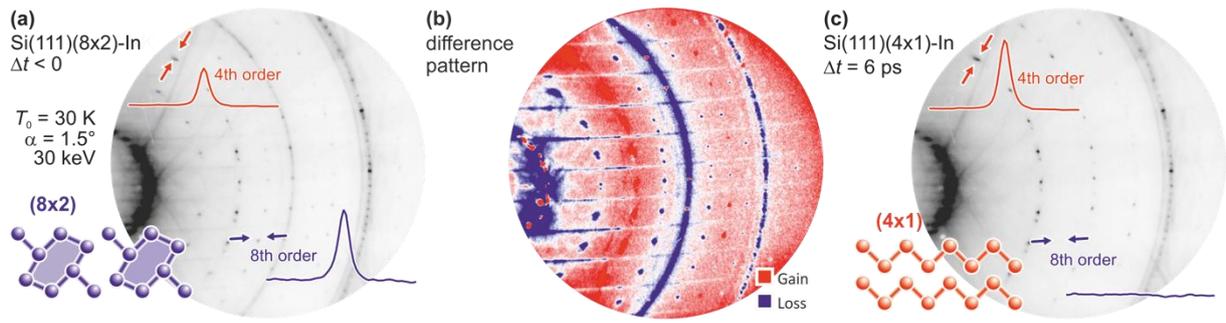

**FIG. 25.** RHEED patterns for clarity shown in inverted intensity representation (bright spots are shown in dark, background in bright) at 30 K prior and after optical excitation through a fs laser pulse. (a) Pattern exhibiting (8×2) ground state. Spot profiles of a 4-fold and an 8-fold spot are shown in red and blue, respectively. (c) The pattern 6 ps after excitation has changed to (4×1). All (8×2) spots and 2-fold streaks are disappeared, as evident from the changes in spot profile, indicating the structural transition. (b) The difference pattern in false color representation exhibits systematic changes: all (4×1) spots gain intensity (red) while the (8×2) and 2-fold streaks disappeared (blue).

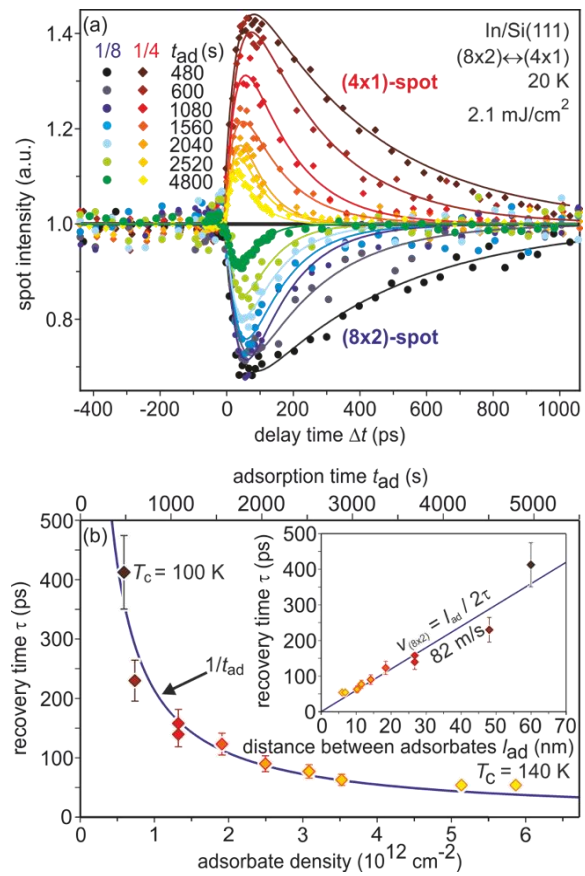

**FIG. 26.** Recovery of the (8×2) ground state. (a) The recovery of the (8×2) ground state strongly depends on adsorption from the residual gas. With increasing adsorbate density, the recovery time constant τ changes from τ = 415 ps for the first experiment at $t_{ad}$ = 480 s (dark red data points) to τ = 54 ps at $t_{ad}$ = 4800 s (light yellow data points). (b) Time constant t for the recovery of the (8×2) reconstruction as a function of adsorbate density. The solid line describes a $1/t_{ad}$ behavior. From the slope in the inset we derive a velocity of the propagating phase front of $v_{(8×2)}$ = 82 m/s.

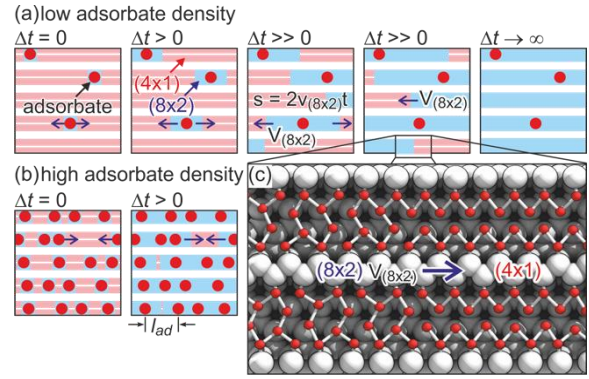

**FIG. 27.** Propagation of the phase front of the (8×2) ground state. (a+b) Adsorbates with a mean separation $l_{ad}$ act as seeds (red dots). $v_{(8×2)}$ is the velocity of the propagating phase front. Low (a) and high (b) adsorbate densities are shown. (c) A snapshot from the ab initio molecular dynamics simulations depicts the transition from the metastable (4×1) phase to the (8×2) ground state.

The transition back to the (8×2) ground state may also be facilitated by condensation nuclei in form of the omnipresent steps on the Si(111) substrate [117]. With the knowledge of the mean terrace width between two atomic steps $<\Gamma> = 350$ nm and the time constant of recovery to the ground state $\tau_{rec} = 3$ ns a speed of the 1D-recovery front of 112 m/s² was determined experimentally [115], which agrees with the above determined value of $v_{(8×2)} = 82$ m/s.

### 3.5.3 Melting of a CDW at the quantum limit

The initial dynamics of this optically driven structural transition was followed in the time domain through the transient intensity changes of RHEED spots as function of pump-probe delay $\Delta t$ as is shown in Fig. 28. Figure 28(a) shows that the 8$^{th}$-order diffraction intensity (indicative for the ground state) is quenched in less than 1 ps. Owing to the much higher signal-to-noise ratio as compared to the 8$^{th}$-order spots, the dynamics of the more intense (00) spot was analyzed which follows the same trend as the 8$^{th}$-order spots. The (00) spot decreases with a time constant of $\tau_{trans} = 350$ fs for a laser fluence of $\Phi = 6.7$ mJ / cm² as shown in Fig. 28(b), i.e., the structural transition is completed in only 700 fs. No oscillatory signatures of the optical phonons connected to the periodic lattice distortion are observed, in contrast to studies on other CDW materials [107,118-120]. This structural transition from the initial insulating (8×2) state to the final metallic (4×1) state is driven by transient changes of the ground state PES which is sketched in Fig. 29(a). Photo excitation of the electron system leads to a depopulation of those states at the top of the surface state conduction band which are responsible for the energy gain through the Peierls distortion as sketched in Fig. 29(b). This results in a transient change of the energy landscape, as is sketched in Fig. 29(c) for $\Delta t = 0.3$ ps. Inevitably, the system undergoes a strongly accelerated displacive structural transition to the minimum of the transient energy landscape. The experimentally determined value of $\tau_{exc} = 350$ fs is about 1/4 of the periods of the equilibrium rotational and shear modes, $T_{rot} = 1.2$ ps and $T_{shear} = 1.8$ ps, respectively [114]. The transition from (8×2) state to the excited (4×1) state is completed after 0.7 ps. The temporal fine structure of one of the (4×1) diffraction spots (Figs. 28(c) and 28(d)) is determined by two opposing trends. First, the initial increase within less than 1 ps is due to the structure factor enhancement of the (4×1) phase reflecting the change of atomic position. Second, the subsequent decrease in intensity is explained by the Debye-Waller effect and results from the excitation of incoherent surface vibrations [25]. This leads to a transient minimum at 6 ps, which is confirmed by the rise of the thermal diffuse background and its temporal evolution (grey circles). We find time constants of 2.2 ps and 30 ps for heating and cooling of the indium atoms, respectively. This situation is sketched in Figs. 29(d) and 29(e) for $\Delta t = 6$ ps and $\Delta t > 100$ ps. From the stationary Debye-Waller behavior of the high temperature (4×1) phase and its extrapolation to lower temperatures we determined the maximum transient temperature $T_{max} = T_0 + \Delta T_{max} = 30$ K $+ 80$ K $= 110$ K at $\Delta t = 6$ ps which is well below $T_c = 130$ K. We therefore conclude that the structural transition occurs with $\tau_{exc} = 350$ fs, well before the initial excitation has thermalized at 6 ps, and thus is not thermally driven.

We observe a threshold fluence of 1 mJ / cm$^2$ below which the system shows some transient response by does not make it into the excited (4×1) state. The excitation of the characteristic shear and rotational phonon modes at a frequency of 0.82 THz and 0.54 THz has been indirectly observed in a pump-pump-probe experiment addressing coherent control of the phase transition.[121] The resulting amplitudon mode is sketched in Fig. 30(b).

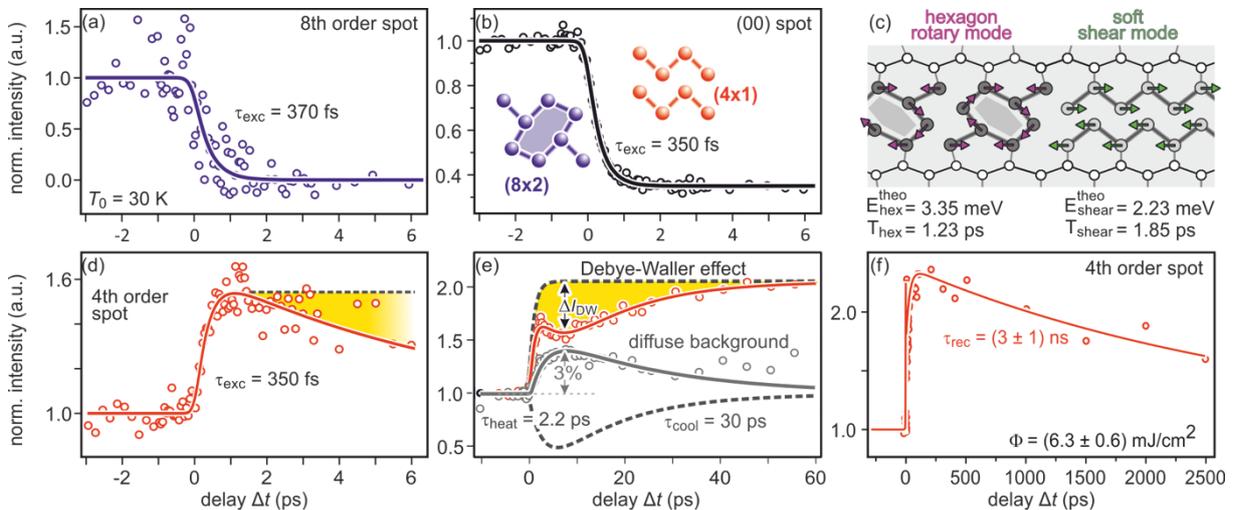

**FIG. 28.** Time evolution of the diffraction intensities following the fs-photoexcitation as a function of pump–probe delay $\Delta t$. Solid lines are (exponential) fits to the data. (a) The transient intensity of an (8×2) spot at a laser fluence of $\Phi = 6.7$ mJ/cm$^2$ vanish at a rate of $\tau_{exc} = 370$ fs to the background level. (b) Transient intensity of the (00) spot reflecting the structural transition from (8×2) to (4×1) state at a rate of $\tau_{exc} = 350$ fs. (c) Characteristic hexagon rotary and soft shear phonon modes facilitating the transition. (d+e) Intensity of a 4$^{th}$-order spot and the thermal diffuse background at a laser fluence of $\Phi = 6.7$ mJ/cm$^2$. The transient dip in the intensity of the 4$^{th}$-order spot $\Delta I_{DBW}$ (yellow shaded area) at $\Delta t = 6$ ps indicates surface heating by $\Delta T = 80$ K, which coincides with the increase in background intensity. The 4$^{th}$-order spot intensity is described (solid red curve) by the superposition of the two dashed lines, representing incoherent thermal motion (heating and subsequent cooling with time constants of 2.2 ps and 30 ps, respectively) and the structural transition with $\tau_{exc} = 350$ fs. (f) metastable state for long timescales. The supercooled (4×1) state recovers slowly on a 3 ns timescale.

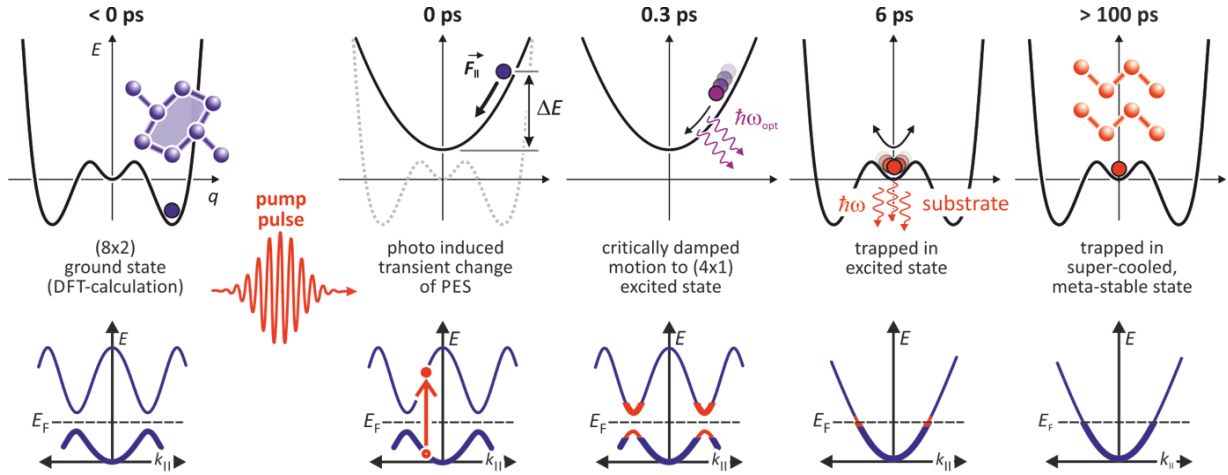

**FIG. 29.** Principle sketch of transient changes of potential energy surface (upper row) and simplified bandstructure (lower row) as function of time delay $\Delta t$. (a) ground state prior to excitation. (b) Photo excitation, generation of electron hole pair, excitation of electron system, transient change of PES. (c) Accelerated displacive structural transition, critically damped motion due to effective energy dissipation to manifold of surface phonon modes. (d) System is trapped in excited high temperature state, electron and lattice system are thermalized. (e) Ground state PES, system trapped in metastable, supercooled state, energy barrier hinders immediate recovery of ground state for nanoseconds.

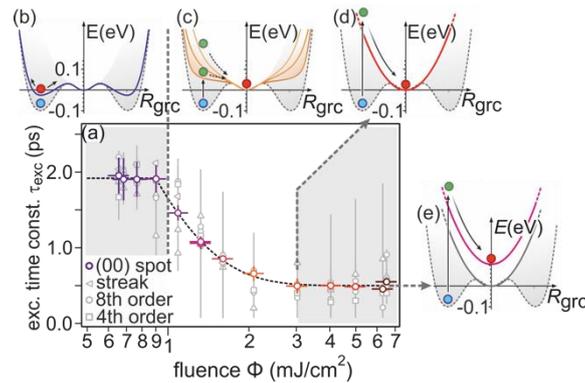

**FIG. 30.** (a) Fluence dependence of excitation time constant $\tau_{exc}$ of driven structural transition. Below $\Phi = 0.9$ mJ/cm$^2$ the (8×2) state is not driven into the excited (4×1) state. The ground state exhibits excitation of the CDW as sketched in (b). For the intermediate regime $0.9$ mJ/cm$^2 < \Phi < 3$ mJ/cm$^2$ the accelerated displacive structural transition into the excited (4×1) state takes place. The slope of the transient PES increases, i.e., speeding up the transition as is sketched in (c). The transition speed saturates for $\Phi \geq 3$ mJ/cm$^2$. The slope of the transient PES is maximum as sketched in (d) and (e).

For an incidence fluence larger than 1 mJ/cm$^2$ the potential energy surface of the (8×2) ground state is transiently lifted above the barrier between the two states as sketched in Fig. 30(c). The surface system undergoes an accelerated transition to the excited (4×1) state in a displacive excitation scenario. Naturally, with increasing laser fluence, i.e., increasing excitation density, the slope of the PES becomes steeper and steeper, resulting in a faster transition to the excited state, as predicted by theory. The transition time constant saturates for incident fluences of 3 mJ/cm$^2$ or more as depicted in Fig. 30(d). This behavior nicely confirms the theoretical predictions where higher excitations densities only shift the PES without increasing its gradient as predicted by theory, see Fig. 31(a). The observed sharp transition from the initial (8×2) state to the final (4×1) state, without any sign of damped oscillatory behavior, is explained through fast mode conversion, dephasing and strong damping of the two characteristic rotary and shear phonon modes [122,123]. Figure 31(d) shows that the initially excited rotary and shear modes rapidly transfer their energy to a manifold of modes at the surface of the Si substrate. The experimentally determined asymptotic value of $\tau_{trans} = 350 \pm 10$ fs is about 1/4 of the periods of the equilibrium rotational and shear modes, $T_{rot} = 1.2$ ps and $T_{shear} = 1.8$ ps [104], respectively. Consequently, the transition proceeds in the regime of critical damping and cannot be faster than in this limit. All of the surface indium atoms move in a spatially coherent manner: the system undergoes the structural transition in the quantum limit: "Quantum limit is the non-statistical regime of rates in which the nuclear motion is directed and deterministic on the shortest scales of length (0.1–1 nm) and time ($10^{-13}$ to $10^{-12}$ s)" [19]. Our results demonstrate that structural transitions at surfaces can be driven as fast as those in bulk materials in the non-thermal regime.

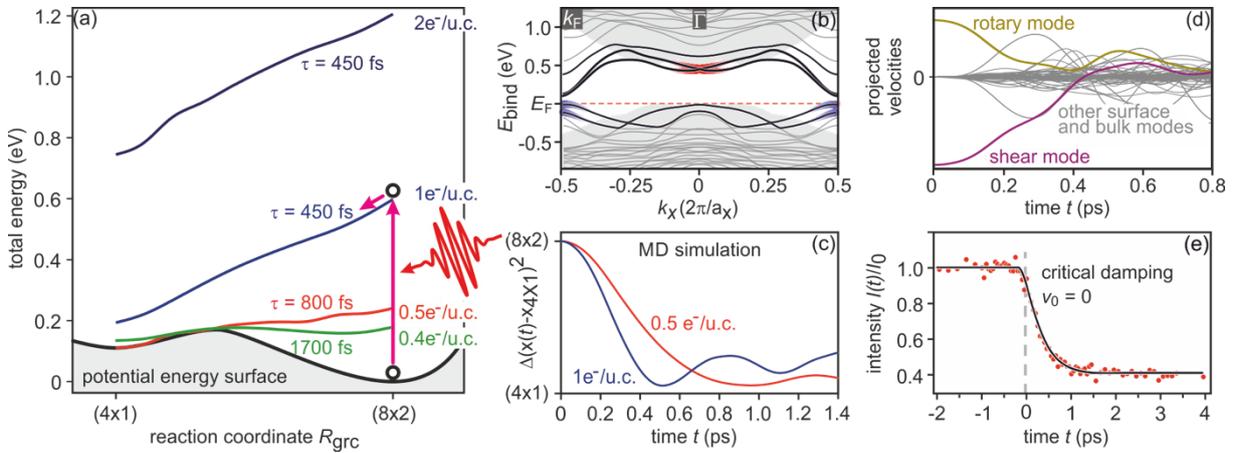

**FIG. 31.** Potential energy surfaces, electronic surface states, molecular dynamics. (a) Calculated potential energy surfaces for the ground state (black) and various excited configurations along the (8×2) → (4×1) minimum-energy path, i.e., along the generalized reaction coordinate $R_{grc}$. The open circles and purple and red arrows indicate excitation of the (8×2) phase. (b) Calculated electronic bands of the Si(111)(8×2)–In surface. Here, $K_x$ and $a_x$ are the reciprocal- and real-space lattice vectors in the wire direction; $E_{bind}$ is the electron binding energy relative to the valence band maximum in silicon. The electron occupation of the blue and red shaded surface bands (black lines) is vital for the phase transition. Grey shaded areas show projected silicon bulk bands. (c) Time evolution of the structural deviation from the (4×1) state, obtained from ab initio molecular dynamics simulations within the adiabatic approximation, for two excited configurations. Here, $X(t)$ and $X(4\times 1)$ denote the atomic coordinates of the In atoms, during the molecular dynamics calculation and for the high-temperature phase, respectively. (d) Transient atomic velocities projected onto vibrational eigenmodes. As evident, the rotary and shear modes rapidly transfer their energy to other modes. (e) The transient intensity $I(t)/I_0$ of the (00)-spot is well fitted by the behavior expected for critical damping (solid black line). With permission from [122].

The optically induced Si(111)(8×2)–In CDW melting relies on transient changes in the PES that arise from the population of very specific electronic states. These directly couple to the two characteristic rotational and shear vibrational lattice modes that drive the structural transition. This melting mechanism is similar to the structural bottleneck mechanism in layered bulk materials [124], which is surprising because the surface system differs from bulk CDWs in one key way. Bulk CDWs are formed within layers or chains that only weakly couple to the environment, and so the signatures of low-dimensional physics remain intact. In contrast, the surface CDW analyzed here is characterized by In–In and In–Si bonds that are stronger than In–In bulk bonds. Despite this strong interaction within the surface and between surface and substrate, the Peierls instability remains. The substrate serves as a skeleton that anchors the indium atoms, but with sufficient freedom to adopt different lateral positions. The strong coupling between substrate and adsorbate facilitates the sub-picosecond structural response by dephasing and damping the characteristic phonons after the structural transition. The structural transition of CDW melting at the Si(111)(8×2)–In surface therefore proceeds in a non-thermal regime in a limit of critical damping of the atomic motion. The CDW interaction with the surface enables the transition timescale to be controlled via the coupling strength of surface atoms to the environment, and opens up possibilities for using femtosecond switching to control and steer energy and matter [125].

**4 Conclusions**

Ultra-fast time resolved RHEED is a versatile tool to study relaxation processes of excited surface systems on a picosecond to femtosecond time scale. In this review, we have demonstrated that possible applications range from heat transport in nanostructures via mode conversion in adsorbate layers to the non-equilibrium dynamics of driven phase transitions. Whenever structural dynamics is considered, ultrafast RHEED will provide new and often unforeseen insights into the non-equilibrium processes at surfaces.
Improved temporal resolution was achieved through the implementation of a tilted pulse front scheme to compensate the velocity mismatch between probing 30 keV electrons and pumping laser pulse. Ultimately an unprecedented temporal response of the entire experiment of 330 fs full width at half maximum of the temporal instrumental response function was obtained.

**Acknowledgements**

We would like to thank M. Aeschlimann, B. Siwick, D. Miller, G. Sciaini, F.J. Meyer zu Heringdorf, P. Baum, D. von der Linde, A. Lücke, W.G. Schmidt, U. Gerstmann, K. Sokolowski-Tinten, M. Ligges, and U. Bovensiepen for many fruitful discussions. Financial support from the Deutsche Forschungsgemeinschaft through the SFB 616 "Energy Dissipation at Surfaces" and SFB 1242 "Non-Equilibrium Dynamics of Condensed Matter in the time domain" (Project-ID 278162697) is gratefully acknowledged.

**References**

[1] C. Davisson and L. H. Germer, "Diffraction of Electrons by a Crystal of Nickel," Phys. Rev. **30**, 705 (1927).

[2] K. Heinz and L. Hammer, "Surface Crystallography by Low Energy Electron Diffraction," Z. Kristallographie **213**, 615 (1998).

[3]     H. E. ElsayedAli and G. A. Mourou, "Picosecond reflection high-energy electron diffraction," Appl. Phys. Lett. **52**, 103 (1988).

[4]     H. E. ElsayedAli and J. W. Herman, "Ultrahigh vacuum picosecond laser-driven electron diffraction system," Rev. Sci. Instrum. **61**, 1636 (1990).

[5]     M. Aeschlimann, E. Hull, J. Cao, C. A. Schmuttenmaer, L. G. Jahn, Y. Gao, H. E. Elsayed-Ali, D. A. Mantell, M. R. Scheinfein, "A picosecond electron gun for surface analysis," Rev. Sci. Instrum. **66**, 1000 (1995).

[6]     B. Krenzer, A. Hanisch-Blicharski, P. Schneider, Th. Payer, S. Möllenbeck, O. Osmani, M. Kammler, R. Meyer and M. Horn-von Hoegen, "Phonon confinement effects in ultra-thin, epitaxial Bismuth-films on Silicon studied by time-resolved electron diffraction, " Phys. Rev. B **80**, 024307 (2009).

[7]     P. Kury, R. Hild, D. Thien, H.-L. Günter, F.-J. Meyer zu Heringdorf, and M. Horn-von Hoegen, "Compact and transferable threefold evaporator for molecular beam epitaxy in ultrahigh vacuum," Rev. Sci. Instrum. **76**, 083906 (2005).

[8]     B. Hafke, T. Witte, C. Brand, M. Horn-von Hoegen, Th. Duden, "Pulsed electron gun for ultrafast time-resolved electron diraction at surfaces with femtosecond temporal resolution and high coherence length," Revi. Sci. Inst. **90**, 045119 (2019).

[9]     Gerard Mourou and Steve Williamson, "Picosecond electron diffraction," Appl. Phys. Lett. **41**, 44 (1982).

[10]    A. Janzen, B. Krenzer, P. Zhou, D. von der Linde, and M. Horn-von Hoegen, "Ultra-fast electron diffraction at surfaces after laser excitation," Surf. Sci. **600**, 4094 (2006).

[11]    A. Janzen, B. Krenzer, O. Heinz, P. Zhou, D. Thien, A. Hanisch, F.-J., Meyer zu Heringdorf, D. von der Linde, M. Horn-von Hoegen, "A pulsed electron gun for ultra-fast electron diffraction at surfaces," Rev. Sci. Inst. **78**, 13906 (2007).

[12]    M. Merano, S. Collin, P. Renucci, M. Gatri, S. Sonderegger, A. Crottini, J. D. Ganière, and B. Deveaud, "High brightness picosecond electron gun," Rev. Sci. Instrum. **76**, 085108 (2005).

[13]    B. J. Siwick, J. R. Dwyer, R. E. Jordan, R. J. D. Miller, "Ultra-fast electron optics: Propagation dynamics of femtosecond electron packets," J. Appl. Phys. **92**, 1643 (2002).

[14]    P. Baum, D.-S. Yang, and A. H. Zewail, "4D visualization of transitional structures in phase transformations by electron diffraction," Science **318**, 788 (2007).

[15]    P. Baum and A. H. Zewail, "Breaking resolution limits in ultrafast electron diffraction and microscopy," Proc. Natl. Acad. Sci. U. S. A. **103**, 16105 (2006).

[16]    P. Zhou, C. Streubühr, A. Kalus, T. Frigge, S. Wall, A. Hanisch-Blicharski, M. Kammler, M. Ligges, U. Bovensiepen, D. von der Linde, and M. Horn-von Hoegen, „Ultrafast time resolved reflection high energy electron diffraction with tilted pump pulse fronts," EPJ Web Conf. **41**, 10016 (2013).

[17]    W. Braun, Applied RHEED, "Reflection high-energy electron diffraction during crystal growth," Springer Tracts in Modern Physics **154**, Springer Verlag (1999).

[18]    M. Henzler and W. Göpel, „Oberflächenphysik des Festkörpers," Teubner Studienbücher Physik, Stuttgart (1991).

[19]    Chong-Yu Ruan, Franco Vigliotti, Vladimir A. Lobastov, Songye Chen, and Ahmed H. Zewail, "Ultrafast electron crystallography: Transient structures of molecules, surfaces, and phase transitions," PNAS **101**, 1123 (2004).

[20]    S. Chen, M. T. Seidel, and A. H. Zewail, "Atomic-Scale Dynamical Structures of Fatty Acid Bilayers Observed by Ultra-fast Electron Crystallography," PNAS **102**, 8854 (2005).

[21]    H. Park and J. M. Zuo, "Direct measurement of transient electric fields induced by ultra-fast pulsed laser irradiation of silicon," Appl. Phys. Lett. **94**, 251103 (2009).

[22]    H. Park and J. M. Zuo, "Ultra-fast Electron Diffraction and Microscopy: Why We Should Know About Transient Electric Fields and Where They Come From," Microscopy and Microanalysis **16**, 496 (2010).

[23] C. Ropers, D. R. Solli, C. P. Schulz, C. Lienau, and T. Elsaesser, "Localized Multiphoton Emission of Femtosecond Electron Pulses from Metal Nanotips," Phys. Rev. Lett. **98**, 043907 (2007).

[24] P. Debye, „Interferenz von Röntgenstrahlen und Wärmebewegung," Ann. d. Phys. **348**, 49 (1913).

[25] I. Waller, „Zur Frage der Einwirkung der Wärmebewegung auf die Interferenz von Röntgenstrahlen," Zeitschrift für Physik **17**, 398 (1923).

[26] E. T. Swartz, R. O. Pohl, "Thermal Resistance at Interfaces," Appl. Phys. Lett. **51**, 2200 (1987).

[27] E. T. Swartz and R. O. Pohl, "Thermal boundary resistance," Rev. Mod. Phys. **61**, 605 (1989).

[28] R. J. Stoner and H. J. Maris, "Kapitza conductance and heat flow between solids at temperatures from 50 to 300 K," Phys. Rev. B **48**, 16373 (1993).

[29] B. Krenzer, A. Janzen, P. Zhou, D. von der Linde, and M. Horn-von Hoegen, "Thermal Boundary Conductance in Heterostructures Studied by Ultra-fast Electron Diffraction," New J. Phys. **8**, 190 (2006)

[30] A. Hanisch, B. Krenzer, T. Pelka, S. Möllenbeck, and M. Horn-von Hoegen. „Thermal response of epitaxial thin Bi films on Si(001) upon femtosecond laser excitation studied by ultra-fast electron diffraction," Phys. Rev. B **77**, 125410 (2008).

[31] T. Frigge, B. Hafke, V. Tinnemann, B. Krenzer, and M. Horn-von Hoegen," Nanoscale heat transport from Ge hut, dome, and relaxed clusters on Si(001) measured by ultrafast electron diffraction," Appl. Phys. Lett. **106**, 053108 (2015).

[32] C. A. Paddock and G. L. Eesley, "Transient thermoreflectance from thin metal films," J. Appl. Phys. **60**, 285 (1986).

[33] C. T. D. A. Young and J. Tauc, "Phonon Scattering in Condensed Matter,"edited by A. C. Anderson and J. P. Wolfe (Springer, Berlin, 1986).

[34] E. Bauer, „Phänomenologische Theorie der Kristallabscheidung an Oberflächen II," Z. für Kristallogr. Cryst. Mater. **110,** 395 (1958).

[35] Y.-W. Mo, D. E. Savage, B. S. Swartzentruber, and M. G. Lagally "Kinetic pathway in Stranski-Krastanov growth of Ge on Si(001)," Phys. Rev. Lett. **65**, 1020 (1990).

[36] F. Montalenti, P. Raiteri, D. Migas, H. Von Känel, A. Rastelli, C. Manzano, G. Costantini, U. Denker, O. Schmidt, K. Kern and Leo Miglio," Atomic-Scale Pathway of the Pyramid-to-Dome Transition during Ge Growth on Si(001)," Phys. Rev. Lett. **93**, 216102 (2004).

[37] I. Goldfarb, P. T. Hayden, J. H. G. Owen, and G. A. D. Briggs, "Competing growth mechanisms of Ge/Si(001) coherent clusters," Phys. Rev. B **56**, 10459 (1997).

[38] M. Kästner and B. Voigtländer, „Kinetically Self-Limiting Growth of Ge Islands on Si(001)," Phys. Rev. Lett. **82**, 2745 (1999).

[39] G. Medeiros-Ribeiro, A. M. Bratkovski, T. I. Kamins, D. A. A. Ohlberg, and R. S. Williams, "Shape Transition of Germanium Nanocrystals on a Silicon (001) Surface from Pyramids to Domes," Science **279**, 353 (1998).

[40] F. M. Ross, R. M. Tromp, and M. C. Reuter, "Transition States Between Pyramids and Domes During Ge/Si Island Growth," Science **286**, 1931 (1999).

[41] M. Horn-von Hoegen, B. H. Müller, T. Grabosch, and P. Kury, " Strain relief during Ge hut cluster formation on Si(001) studied by high-resolution LEED and surface-stress-induced optical deflection," Phys. Rev. B **70**, 235313 (2004).

[42] A. J. Schell-Sorokin and R. M. Tromp, "Mechanical stresses in (sub)monolayer epitaxial films," Phys. Rev. Lett. **64**, 1039 (1990).

[43] G. Wedler, J. Walz, T. Hesjedal, E. Chilla, and R. Koch, "Stress and Relief of Misfit Strain of Ge/Si(001)," Phys. Rev. Lett. **80**, 2382 (1998).

[44] A. J. Steinfort, P. M. L. O. Scholte, A. Ettema, F. Tuinstra, M. Nielsen, E. Landemark, D.-M. Smilgies, R. Feidenhans'l, G. Falkenberg, L. Seehofer, and R. L. Johnson, „Strain in Nanoscale Germanium Hut Clusters on Si(001) Studied by X-Ray Diffraction," Phys. Rev. Lett. **77**, 2009 (1996).

[45]　　G. Medeiros-Ribeiro, T. Kamins, D. Ohlberg, and R. S. Williams, "Annealing of Ge nanocrystals on Si(001) at 550°C: Metastability of huts and the stability of pyramids and domes," Phys. Rev. B **58**, 3533 (1998).

[46]　　C.-P. Liu, J. M. Gibson, D. G. Cahill, T. I. Kamins, D. P. Basile, and R. S. Williams, "Strain Evolution in Coherent Ge/Si Islands," Phys. Rev. Lett. **84**, 1958 (2000).

[47]　　F. Ross, J. Tersoff, and R. Tromp, "Coarsening of Self-Assembled Ge Quantum Dots on Si(001)," Phys. Rev. Lett. **80**, 984 (1998).

[48]　　A. Nikiforov, V. Cherepanov, O. Pchelyakov, A. Dvurechenskii, and A. Yakimov, In situ RHEED control of self-organized Ge quantum dots," Thin Solid Films **380**, 158 (2000).

[49]　　D. J. Eaglesham and M. Cerullo, "Dislocation-free Stranski-Krastanow growth of Ge on Si(100)," Phys. Rev. Lett. **64**, 1943 (1990).

[50]　　S. Chaparro, Y. Zhang, J. Drucker, D. Chandrasekhar, and D. J. Smith, "Evolution of Ge/Si(100) islands: Island size and temperature dependence," J. Appl. Phys. **87**, 2245 (2000).

[51]　　C.-Y. Ruan, V. A. Lobastov, F. Vigliotti, S. Chen, and A. H. Zewail, "Ultrafast Electron Crystallography of Interfacial Water," Science **304**, 80 (2004).

[52]　　C. E. Aumann, Y.-W. Mo, and M. G. Lagally, "Diffraction determination of the structure of metastable three-dimensional crystals of Ge grown on Si(001)," Appl. Phys. Lett. **59**, 1061 (1991).

[53]　　J. M. Ziman, "Principles of the Theory of Solids," Cambridge University Press, 64 (1979).

[54]　　K. Sokolowski-Tinten, U. Shymanovich, M. Nicoul, J. Blums, A. Tarasevitch, M. Horn-von Hoegen, D. von der Linde, A. Morak, and T. Wietler, „Energy relaxation and anomalies in the thermo-acoustic response of femtosecond laser-excited Germanium," in Ultrafast Phenomena XV, Springer Series in Chemical Physics **88**, 597, edited by Weiner, A.M.; Miller, R. J. D. (Eds.) (2006).

[55]　　A. Hanisch-Blicharski, A. Janzen, B. Krenzer, S. Wall, F. Klasing, A. Kalus, T. Frigge, M. Kammler, and M. H. von Hoegen, „Ultra-fast electron diffraction at surfaces: From nanoscale heat transport to driven phase transitions," Ultramicroscopy **127**, 2 (2013).

[56]　　A. B. Shick, J. B. Ketterson, D. L. Novikov, and A. J. Freeman, "Electronic structure, phase stability, and semimetal-semiconductor transitions in bi," Phys. Rev. B **60**, 15484 (1999).

[57]　　H. J. Zeiger and et al., "Theory for displacive excitation of coherent phonons," Phys. Rev. B **45**, 768 (1992).

[58]　　T. K. Cheng, J. Vidal, H. J. Zeiger, G. Dresselhaus, M. S. Dresselhaus, and E. P. lppen, "Mechanism for displacive excitation of coherent phonons in Sb, Bi, Te, and $Ti_2O_3$," Appl. Phys. Lett. **59,** 1923 (1991).

[59]　　Muneaki Hase, Kohji Mizoguchi, Hiroshi Harima, Shinichi Nakashima, and Kiyomi Sakai, "Dynamics of coherent phonons in bismuth generated by ultrashort laser pulses," Phys. Rev. B **58,** 5448 (1998).

[60]　　M. Hase, M. Kitajima, S. Nakashima, and K. Mizoguchi, "Dynamics of coherent anharmonic. phonons in bismuth using high density photoexcitation," Phys. Rev. Lett. **88**, 067401 (2002).

[61]　　K. Sokolowski-Tinten, C. Blome, J. Blums, A. Cavalleri, C. Dietrich, A. Tarasevitch, I. Uschmann, E. Förster, M. Kammler, M. Horn-von Hoegen, and D. von der Linde, "Femtosecond x-ray measurement of coherent lattice vibrations near the lindemann stability limit," Nature **422**, 287 (2003).

[62]　　S. L. Johnson, P. Beaud, C. J. Milne, F. S. Krasniqi, E. S. Zijlstra, M. E. Garcia, M. Kaiser, D. Grolimund, R. Abela, and G. Ingold, "Nanoscale depth-resolved coherent femtosecond motion in laser-excited bismuth," Phys. Rev. Lett. **100**, 155501 (2008).

[63]　　S. L. Johnson, P. Beaud, E. Vorobeva, C. J. Milne, E. D. Murray, S. Fahy, and G. Ingold, "Nanoequilibrium phonon dynamics studied by grazing-incidence femtosecond x-ray crystallography," Acta Cryst. **A66**, 157 (2010).

[64]　　G. Sciaini, M. Harb, S. G. Kruglik, T. Payer, C. T. Hebeisen, F.-J. Meyer zu Heringdorf, M. Yamaguchi, M. Horn-von Hoegen, R. Ernstorfer, and R. J. D. Miller, "Electronic acceleration of atomic motions and disordering in bismuth," Nature **458**, 56 (2009).

[65]   D. M. Fritz, D. A. Reis, B. Adams, R. A. Akre, J. Arthur, C. Blome, P. H. Bucksbaum, A. L. Cavalieri, S. Engemann, S. Fahy, R. W. Falcone, P. H. Fuoss, K. J. Gaffney, M. J. George, J. Hajdu, M. P. Hertlein, P. B. Hillyard, M. Horn-von Hoegen, M. Kammler, J.Kaspar, R. Kienberger, P. Krejcik, S. H. Lee, A. M. Lindenberg, B. McFarland, D. Meyer, T. Montagne, D. Murray, A. J. Nelson, M. Nicoul, R. Pahl, J. Rudati, H. Schlarb, D. P. Siddons, K. Sokolowski-Tinten, Th. Tschentscher, D. von der Linde, and J. B. Hastings, "Ultrafast bond softening in bismuth: Mapping a solid's interatomic potential with x-rays," Science **315**, 633 (2007).

[66]   M. F. DeCamp, D. A. Reis, P. H. Bucksbaum, and R. Merlin, "Dynamics and coherent control of high-amplitude optical phonons in bismuth," Phys. Rev. B **64**, 092301 (2001).

[67]   S. L. Johnson, P. Beaud, E. Vorobeva, C. J. Milne, E. D. Murray, S. Fahy, and G. Ingold, "Directly observing squeezed phonon states with femtosecond x-ray diffraction," Phys. Rev. Lett. **102**, 175503 (2009).

[68]   A.R. Esmail and H. E. Elsayed-Ali, "Anisotropic response of nanosized bismuth films upon femtosecond laser excitation monitored by ultrafast electron diffraction," Appl. Phys. Lett. **99**,161905 (2011).

[69]   C. Streubühr, A. Kalus, P. Zhou, M. Ligges, A. Hanisch-Blicharski, M. Kammler, U. Bovensiepen, M. Horn-von Hoegen, and D. von der Linde, "Comparing ultrafast surface and bulk heating using time-resolved electron diffraction," Appl. Phys. Lett. **104**, 161611 (2014).

[70]   K. Sokolowski-Tinten, R. Li, A. Reid, S. Weathersby, F. Quirin, T. Chase, R. Coffee, J. Corbett, A. Fry, N. Hartmann, C. Hast, R. Hettel, M. Horn-von Hoegen, D. Janoschka, J. Lewandowski, M. Ligges, F. Meyer zu Heringdorf, X. Shen, T. Vecchione, C. Witt, J. Wu, H. Dürr, andX. Wang, "Thickness-dependent electron-lattice equilibration in laser-excited thin bismuth films," New J. Phys **17**, 113047 (2015).

[71]   H. Hase, K. Ishioka, M. Kitajima, S. Hishita, and K. Ushida, "Dephasing of coherent THz phonons in bismuth studied by femtosecond pump probe technique," Appl. Surf. Sci. **197-198**, 710 (2002).

[72]   P. Fischer, I. Sosnowska, and M. Szymanski, "Debye-waller factor and thermal expansion of arsenic, antimony and bismuth," J. Phys. C **11**, 1043 (1978).

[73]   M. Kammler and M. Horn-von Hoegen, "Low energy electron diffraction of epitaxial growth of bismuth on si(111)," Surf. Sci. **576**, 56 (2005).

[74]   A. Hanisch-Blicharski, „Ultraschnelle Elektronenbeugung an Oberflächen zur Untersuchung des ballistischen Wärmetransports in nanoskaligen Heterosystemen," Ph.D. thesis, University of Duisburg-Essen (2009).

[75]   H. Boersch, G. Jeschke, and D. Willasch, "Temperature dependence of the electron diffraction intensities (1120) of bismuth in the case of anomalous electron transmission," Phys. Lett. A **29**, 493 (1969).

[76]   G. Jeschke and D. Willasch, "Temperaturabhängigkeit der anomalen elektronenabsorption von wismut-einkristallen," Zeitschrift für Physik **238**, 421 (1970).

[77]   Manuel Ligges, Ivan Rajkovic, Carla Streubühr, Thorsten Brazda, Ping Zhou, Oliver Posth, Christoph Hassel, Günter Dumpich, and Dietrich von der Linde, "Transient (000)-order attenuation effects in ultrafast transmission electron diffraction," J. Appl. Phys. **109**, 063519 (2011).

[78]   V. Tinnemann, C. Streubühr, B. Hafke, A. Kalus, A. Hanisch-Blicharski, M. Ligges, P. Zhou, D. von der Linde, U. Bovensiepen, and M. Horn-von Hoegen. "Ultrafast electron diffraction from a Bi(111) surface: Impulsive lattice excitation and Debye-Waller analysis at large momentum transfer" Struct. Dyn. **6**, 035101 (2019).

[79]   Ph. Hofmann, "The surfaces of bismuth: Structural and electronic properties," Surface Science **81**, 191 (2006).

[80]   É. I. Rashba, "Properties of semiconductors with an extremum loop. 1. cyclotron and combinational resonance in a magnetic field perpendicular to the plane of the loop," Sov. Phys. Solid State **2**, 1190 (1960).

[81]   L. Petersen and P. Hedegard, "A simple tight-binding model of spin-orbit splitting of sp-derived surface states," Surf. Sci. **459**, 49 (2000).

[82] C. R. Ast and H. Höchst, "Fermi surface of bi(111) measured by photoemission spectroscopy," Phys. Rev. Lett. **87**, 177602 (2001).

[83] Yu. M. Koroteev, G. Bihlmayer, J. E. Gayone, E. V. Chulkov, S. Blügel, P. M. Echenique, and Ph. Hofmann, "Strong spin-orbit splitting on bi surfaces," Phys. Rev. Lett. **93**, 046403 (2004).

[84] T. Hirahara, T. Nagao, I. Matsuda, G. Bihlmayer, E. V. Chulkov, Yu. M. Koroteev, P. M. Echenique, M. Saito, and S. Hasegawa, "Role of spin-orbit coupling and hybridization effect in the electronic structure of ultrathin bi films," Phys. Rev. Lett. **97**, 146803 (2006).

[85] J. Faure, J. Mauchain, E. Papalazarou, M. Marsi, D. Boschetto, I. Timrov, N. Vast, Y. Ohtsubo, B. Arnaud, and L. Perfetti, "Direct observation of electron thermalization and electron-phonon coupling in photoexcited bismuth," Phys. Rev. B **88**, 075120 (2013).

[86] M. Shen, W. J. Evans, D. Cahill, and P. Keblinski, "Bonding and pressure-tunable interfacial thermal conductance," Phys. Rev. B **84**, 195432 (2011).

[87] M. Bonn, S. Funk, C. Hess, D. N. Denzler, C. Stamp_, M. Sche_er, M. Wolf, and G. Ertl, „Phonon- Versus Electron-Mediated Desorption and Oxidation of CO on Ru(0001)," Science **285**, 1042 (1999).

[88] J. D. Beckerle, R. R. Cavanagh, M. P. Casassa, E. J. Heilweil, and J. C. Stephenson, "Subpicosecond transient infrared spectroscopy of adsorbates. Vibrational dynamics of CO/Pt(111)," J. Chem. Phys. **95**, 5403 (1991).

[89] K. Laÿ, X. Han, and E. Hasselbrink, "The surprisingly short vibrational lifetime of the internal stretch of CO adsorbed on Si(100)," J. Chem. Phys. **123**, 051102 (2005).

[90] S. Sakong, P. Kratzer, X. Han, T. Balgar, and E. Hasselbrink, „Isotope effects in the vibrational lifetime of hydrogen on germanium(100): Theory and experiment," J. Chem. Phys. **131**, 124502 (2009).

[91] P. Guyot-Sionnest, P. Dumas, Y. J. Chabal, and G. S. Higashi, "Lifetime of an adsorbate-substrate vibration: H on Si(111)," Phys. Rev. Lett. **64**, 2156 (1990).

[92] C. Ulrich, E. Anastassakis, K. Syassen, A. Debernardi, and M. Cardona, "Lifetime of Phonons in Semiconductors under Pressure," Phys. Rev. Lett. **78**, 1283 (1997).

[93] P. Estrup and J. Morrison, "Studies of monolayers of lead and tin on Si(111) surfaces," Surf. Sci **2**, 465 (1964).

[94] T.-L. Chan, C. Z. Wang, M. Hupalo, M. C. Tringides, Z.-Y. Lu, and K. M. Ho, "First-principles studies of structures and stabilities of Pb/Si(111)," Phys. Rev. B **68**, 045410 (2003).

[95] P. Cudazzo, G. Profeta, and A. Continenza, "Low temperature phases of Pb/Si(111) and related surfaces," Surf. Sci. **602**, 747 (2008).

[96] S. Sakong, P. Kratzer, S. Wall, A. Kalus, and M. Horn-von Hoegen, „Mode conversion and long-lived vibrational modes in lead monolayers on silicon (111) after femtosecond laser excitation: A molecular dynamics simulation," Phys. Rev. B **88**, 115419 (2013).

[97] H. W. Yeom, S. Takeda, E. Rotenberg, I. Matsuda, K. Horikoshi, J. Schaefer, C. M. Lee, S. D. Kevan, T. Ohta, T. Nagao, and S. Hasegawa, "Instability and charge density wave of metallic quantum chains on a silicon surface," Phys. Rev. Lett. **82**, 4898 (1999).

[98] C. Kumpf, O. Bunk, J. Zeysing, Y. Su, M. Nielsen, R. Johnson, R. Feidenhans, and K. Bechgaard, "Lowtemperature structure of indium quantum chains on silicon," Phys. Rev. Lett. **85**, 4916 (2000).

[99] K. Sakamoto, H. Ashima, H. W. Yeom, and W. Uchida, "Angle-resolved high-resolution electron-energy-loss study of In-adsorbed Si(111)-(4×1) and-(8×2) surfaces," Phys. Rev. B **62**, 9923 (2000).

[100] H. Yeom, K. Horikoshi, H. Zhang, K. Ono, and R. Uhrberg, "Nature of the broken-symmetry phase of the one-dimensional metallic In/Si (111) surface," Phys. Rev. B **65**, 241307 (2002).

[101] J. R. Ahn, J. H. Byun, H. Koh, E. Rotenberg, S. D. Kevan, and H. W. Yeom, "Mechanism of gap opening in a triple-band Peierls system: In atomic wires on Si," Phys. Rev. Lett. **93**, 106401 (2004).

[102] S. J. Park, H.-W. Yeom, S.-H. Min, D.-H. Park, and I.- W. Lyo, "Direct evidence of the charge ordered phase transition of indium nanowires on si (111)," Phys. Rev. Lett. **93**, 106402 (2004).

[103] J. Guo, G. Lee, and E. W. Plummer, "Intertwined electronic and structural phase transitions in the In/Si (111) interface," Phys. Rev. Lett. **95**, 046102 (2005).

[104] S. Wippermann and W. G. Schmidt, "Entropy explains metal-insulator transition of the Si(111)-In nanowire array," Phys. Rev. Lett. **105**, 126102 (2010).

[105] T. Uchihashi and U. Ramsperger, "Electron conduction through quasi-one-dimensional indium wires on silicon," Appl. Phys. Lett. **80**, 4169 (2002).

[106] T. Tanikawa, I. Matsuda, T. Kanagawa, and S. Hasegawa,"Surface-state electrical conductivity at a metal-insulator transition on silicon," Phys. Rev. Lett. **93**, 016801 (2004).

[107] F. Klasing, T. Frigge, B. Hafke, B. Krenzer, S. Wall, A. Hanisch-Blicharski, and M. Horn-von Hoegen, "Hysteresis proves that the In/Si(111) (8×2) to(4×1) phase transition is first-order," Phys. Rev. B **89**, 121107 (2014).

[108] T. Shibasaki, N. Nagamura, T. Hirahara, and H. Okino, "Phase transition temperatures determined by different experimental methods: Si (111) 4x1-In surface with defects," Phys. Rev. B **81**, 035314 (2010).

[109] S. Wall, B. Krenzer, S. Wippermann, S. Sanna, F. Klasing, A. Hanisch-Blicharski, M. Kammler, W. G. Schmidt, and M. Horn-von Hoegen, "Atomistic picture of charge density wave formation at surfaces," Phys. Rev. Lett. **109**, 186101 (2012).

[110] W. G. Schmidt, S. Wippermann, S. Sanna, M. Babilon, N. J. Vollmers, and U. Gerstmann, "In-Si(111) (4×1)/(8×2) nanowires: Electron transport, entropy, and metal-insulator transition," Phys. Status Solidi B **249**, 343 (2012).

[111] E. Jeckelmann, S. Sanna, W. G. Schmidt, E. Speiser, and N. Esser, "Grand canonical Peierls transition in In/Si(111)," Phys. Rev. B **93**, 241407 (2016).

[112] C. Gonz´alez, J. Ortega, and F. Flores, "Metal-insulator transition in one-dimensional In-chains on Si(111): combination of a soft shear distortion and a double-band Peierls instability," NJP **7**, 100 (2005).

[113] S. Riikonen, A. Ayuela, and D. S´anchez-Portal, "Metal– insulator transition in the In/Si(111) surface," Surf. Sci. **600**, 3821 (2006).

[114] S. Hellmann, T. Rohwer, M. Kalläne, K. Hanff, C. Sohrt, A. Stange, A. Carr, M. Murnane, H. Kapteyn, L. Kipp, et al., "Time-domain classification of charge-densitywave insulators," Nature communications **3**, 1 (2012).

[115] H.-J. Kim and J.-H. Cho, "Driving force of phase transition in indium nanowires on Si(111)," Phys. Rev. Lett. **110**, 116801 (2013).

[116] G. Lee, S.-Y. Yu, H. Shim, W. Lee, and J.-Y. Koo, "Roles of defects induced by hydrogen and oxygen on the structural phase transition of Si(111)4×1-In," Phys. Rev. B **80**, 075411 (2009).

[117] B. Hafke, T. Witte, D. Janoschka, P. Dreher, F.-J. Meyer zu Heringdorf, and M. Horn-von Hoegen, "Condensation of ground state from a supercooled phase in the si (111)-(4×1)→(8×2)-indium atomic wire system," Struct. Dyn. **6**, 045101 (2019).

[118] H. Schäfer, V. V. Kabanov, M. Beyer, K. Biljakovic, and J. Demsar, "Disentanglement of the electronic and lattice parts of the order parameter in a 1D charge density wave system probed by femtosecond spectroscopy," Phys. Rev. Lett. **105**, 066402 (2010).

[119] L. Perfetti, P. Loukakos, M. Lisowski, U. Bovensiepen, H. Berger, S. Biermann, P. Cornaglia, A. Georges, and M. Wolf, "Time evolution of the electronic structure of 1 t- tas 2 through the insulator-metal transition," Phys. Rev. Lett. **97**, 067402 (2006).

[120] L. Rettig, R. Cort´es, J.-H. Chu, I. Fisher, F. Schmitt, R. Moore, Z.-X. Shen, P. Kirchmann, M. Wolf, and U. Bovensiepen, "Persistent order due to transiently enhanced nesting in an electronically excited charge density wave," Nature communications **7**, 1 (2016).

[121] J. G. Horstmann, H. Böckmann, B. Wit, F. Kurtz, G. Storeck, and C. Ropers, "Coherent control of a surface structural phase transition," Nature **583**, 232 (2020).

[122]     T. Frigge, B. Hafke, T. Witte, B. Krenzer, C. Streubühr, A. Samad Syed, V. Mikšić Trontl, I. Avigo, P. Zhou, M. Ligges, D. von der Linde, U. Bovensiepen, M. Horn-von Hoegen, S. Wippermann, A. Lücke, S. Sanna, U. Gerstmann, and W. G. Schmidt, "Optically excited structural transition in atomic wires on surfaces at the quantum limit," Nature **544**, 207 (2017).

[123]     Y.-X. Gu, W.-H. Liu, Z. Wang, S.-S. Li, L.-W. Wang, and J.-W. Luo, "Origin of immediate damping of coherent oscillations in photoinduced charge density wave transition," Phys. Rev. Lett. **130**, 146901 (2023.

[124]     C. Sohrt, A. Stange, M. Bauer, and K. Rossnagel, "How fast can a Peierls–Mott insulator be melted?," Faraday discussions **171**, 243 (2014).

[125]     G. R. Fleming and M. A. Ratner, "Grand challenges in basic energy sciences," Phys. Today **61**, 28 (2008).